\documentclass[ejs]{imsart}

\RequirePackage[OT1]{fontenc}
\usepackage{amsthm,amsmath,amssymb}
\usepackage{graphicx,float,psfrag,epsfig,}
\RequirePackage[numbers]{natbib}
\RequirePackage[colorlinks,citecolor=blue,urlcolor=blue]{hyperref}

\doi{10.1214/154957804100000000}
\pubyear{0000}
\volume{0}
\firstpage{0}
\lastpage{0}

\startlocaldefs

\theoremstyle{definition}
\newtheorem{propo}{Proposition}[section]                                                                                                                                                                                                                                                                                                                                                                                                                                                                                                                                                                                                                                                                                                                                                                                                                                                                                                                                                                                                                                                                                                                                                                                                                                                                                                                                                                                                                                                                                                                                                                                                                                                                                                                                                                                                              
\newtheorem{lemma}[propo]{Lemma}
\newtheorem{definition}[propo]{Definition}
\newtheorem{coro}[propo]{Corollary}
\newtheorem{thm}[propo]{Theorem}

\newtheorem{remark}[propo]{Remark}

\def\hS{\widehat{S}}

\def\tx{\widetilde{x}}

\def\cG{{\cal G}}

\def\hsign{\widehat{\rm{sign}}}
\def\sign{{\rm{sign}}}

\def\dFDP{{\rm{FDP_{dir}}}}

\def\hS{\widehat {S}}

\def\cG{\mathcal{G}}

\def\reals{{\mathbb R}}

\def\eps{{\varepsilon}}
\def\prob{{\mathbb P}}
\def\E{{\mathbb E}}

\def\L0{{L_i}}

\def\de{{\rm d}}
\def\<{\langle}
\def\>{\rangle}
\def\diag{{\rm diag}}

\def\hth{\widehat{\Lambda}}

\def\dth{\widehat{\theta}^{{\rm d}}}

\def\hSigma{\widehat{\Sigma}}

\def\hsigma{\widehat{\sigma}}

\def\supp{{\rm supp}}

\def\F{{\sf F}}
\def\ind{{\mathbb I}}

\def\F{{\sf F}}
\def\normal{{\sf N}}

\def\P{{\mathbb{P}}}

\def\sT{{\sf T}}

\def\id{{\rm I}}

\def\hgamma{\hat{\gamma}}
\def\htau{\hat{\tau}}

\def\sign{{\rm sign}}

\def\v*{v_i}
\def\T*{T_i}

\def\u*{u_i}
\def\F*{F_i}

\def\minimize{{\rm {minimize}}}
\def\subjectto{{\rm {subject\,\,to}}}

\def\dist{\overset{{\rm d}}{\longrightarrow}}

\def\FDR{{\rm FDR}}

\def\FDP{{\rm FDP}}

\def\tT{\widetilde{T}}

\def\htheta{\widehat{\theta}}

\def\power{{\rm Power}}

\def\tx{\tilde{x}}

\def\hth{{\widehat{\theta}}}

\def\diag{{\rm diag}}

\def\FDRdir{{\FDR_{\rm dir}}}
\def\FDPdir{{\FDP_{\rm dir}}}
\def\hsign{\widehat{\sign}}
\def\tZ{\widetilde{Z}}
\def\tT{\widetilde{T}}
\def\pset{{S_{\ge 0}}}
\def\nset{{S_{\le 0}}}

\endlocaldefs

\begin{document}

\begin{frontmatter}

\title{False Discovery Rate Control via Debiased Lasso}
\runtitle{False Discovery Rate Control via Debiased Lasso}

\begin{aug}
\author{\fnms{Adel} \snm{Javanmard}\thanksref{t1}\corref{}\ead[label=e1]{ajavanma@usc.edu}}
\address{Data Sciences and Operations Department, University
of Southern California\\
\printead{e1}}
\author{\fnms{Hamid} \snm{Javadi} \ead[label=e2]{hrhakim@rice.edu}}
\address{Department of Electrical and Computer Engineering, Rice University \\
\printead{e2}}

\thankstext{t1}{A. Javanmard is supported in part by NSF CAREER Award 1844481 and a Google Faculty Research Award. A. Javanmard would also like
to acknowledge the financial support of the Office of the Provost at the University of Southern
California through the Zumberge Fund Individual Grant Program.}

\runauthor{A. Javanmard and H. Javadi}

\affiliation{University of Southern California and Rice University}

\end{aug}

\begin{abstract}

We consider the problem of variable selection in high-dimensional statistical models where the goal is to report a set of variables, out of many predictors $X_1, \dotsc, X_p$, that are relevant to a response of interest. For linear high-dimensional model, where the number of parameters exceeds the number of samples $(p>n)$, we propose a procedure for variables selection and prove that it controls the \emph{directional} false discovery rate (FDR) below a pre-assigned significance level $q\in [0,1]$. We further analyze the statistical power of our framework and show that for designs with subgaussian rows and a common precision matrix $\Omega\in\reals^{p\times p}$, if the minimum nonzero parameter $\theta_{\min}$ satisfies $$\sqrt{n} \theta_{\min} - \sigma \sqrt{2(\max_{i\in [p]}\Omega_{ii})\log\left(\frac{2p}{qs_0}\right)} \to \infty\,,$$ then this procedure achieves asymptotic power one. 

Our framework is built upon the debiasing approach and assumes the standard condition $s_0 = o(\sqrt{n}/(\log p)^2)$, where $s_0$ indicates the number of true positives among the $p$ features. Notably, this framework achieves exact directional FDR control without any assumption on the amplitude of unknown regression parameters, and does not require any knowledge of the distribution of covariates or the noise level. We test our method in synthetic and real data experiments to assess its performance and to corroborate our theoretical results. 
  
\end{abstract}

\begin{keyword}[class=MSC]
\kwd[Primary ]{62F03}
\kwd{62J05}
\kwd{62J07}
\kwd[; secondary ]{62F12}
\end{keyword}

\begin{keyword}
\kwd{Inference in high-dimensional regression}
\kwd{Hypothesis testing}
\kwd{False discovery rate}
\kwd{Model selection}
\kwd{Lasso}
\kwd{Debiased estimator}
\end{keyword}


\tableofcontents

\end{frontmatter}

\section{Introduction}
Living in the era of data deluge, modern datasets are often very fine-grained, including information on a large number of potential explanatory variables. For a given response of interest, we know a priori that a large portion of these variables are irrelevant and would like to select a set of predictors that influence the response. For example, in genome-wide association studies (GWAS), we collect single nucleotide polymorphism (SNP) information across a large number of loci and then aim at finding loci that are related to the trait, while being resilient to false associations. 

The focus of this paper is on high-dimensional regression models where the number of parameters exceeds the sample size. Since such models are over-parameterized, they are prone to overfitting. In addition, high-dimensionality brings noise accumulation and spurious correlations between response and unrelated features, which may lead to wrong statistical inference and false predictions. Model selection is therefore a crucial task in analyzing high-dimensional models. For a successful model selection, we need to assure that most of the selected predictors are indeed relevant. This not only leads to noise reduction and enhances predictions but also offers reproducibility. 

To be concrete in using the term ``reproducibility", we characterize it for statistical inference problem by using the false discovery rate (FDR) criterion, which is the expected fraction of discoveries that are false positives. The notion of FDR has been proposed by the groundbreaking work \cite{benjamini1995controlling} and nowadays is the criterion of choice for statistical inference in large scale hypothesis testing problem. In their work, Benjamini and Hochberg developed a procedure to control FDR under a pre-assigned significance level. It has been shown theoretically that BH procedure controls FDR in some special cases such as independence or positive dependence of tests~\cite{benjamini1995controlling,benjamini2001control}. Since initially proposed, there have been various modifications of BH~\cite{benjamini2001control,Sunetal2015,FanHanGu2012,Wu2008,XieCaiMarisLi2011} and its applications in different domains~\cite{reiner2003identifying,genovese2002thresholding}. 

Importantly, BH procedure (and its modifications) assumes that $p$-values are given as input for all the hypothesis tests. The $p$-values are often calculated using classical formula obtained by using large-sample theory which are theoretically justified only for the classical setting of fixed dimension $p$ and diverging sample size $n$~\cite{van2000asymptotic}. For example, \cite{liu2014phase} considers the setting where $m$ i.i.d random samples of $(X_1, \dotsc, X_p)$ are given and $p$-values are estimated from the Student's t-test statistic. The authors propose a bootstrap calibration method to use with the BH procedure and show that, under weak dependence among observations, it can control the false discovery rate  when the total number of observations $n = mp$ is bigger than $p(\log p)^c$. However, for high-dimensional models obtaining valid $p$-values is highly nontrivial. This is in part due to the fact that fitting high-dimensional model often requires the use of nonlinear and non-explicit estimation procedures and characterizing the distribution of such estimators is extremely challenging.  
In the past couple years, there has been a surge of interest in constructing frequentist $p$-values and confidence intervals for high-dimensional models. A common approach is the fundamental idea of debiasing which was proposed in a series of work~\cite{javanmard2013hypothesis,zhang2014confidence,javanmard2014confidence,van2014asymptotically,javanmard2013nearly,belloni2014inference}. In this approach, starting from a regularized estimator one first constructs a debiased estimator and then makes inference based on the asymptotic normality of low-dimensional functionals
of the debiased estimator. This approach also provides asymptotically valid $p$-values for the null hypotheses of the form $H_{0}: \theta_{0,i} = 0$, where $\theta_{0,i}$ is a fixed single model parameter. However, these $p$-values are correlated and the BH procedure is not guaranteed to control FDR in this case. The modification of BH for general dependency, that scales the significance level by $1/(\log p)$ factor~\cite{benjamini2001control}, also turns out to be overly conservative and leads to a low power. In \cite{belloni2018high}, the authors review the methods for constructing $p$-values in the high dimensional setting, their behavior and limitations, and describe a general set of assumptions under which these $p$-values can be used for inference tasks, such as finding confidence intervals and controlling FDR. In particular, \cite{belloni2018high} extends the result of~\cite{liu2014phase} to the so-called ``Many Approximation Means (MAM)" framework and provide a set of conditions on the dependence among $p$-values such that the Benjamini-Hochberg procedure has the FDR control property.

In this paper, we build upon the debiasing approach and propose a procedure for model selection under the high-dimensional regime, which is guaranteed to have FDR under a pre-assigned level $\alpha\in [0,1]$. 
We call our procedure \emph{FCD} (for ``\emph{FDR Control via Debiasing}") and prove that it controls even a stronger criterion, namely \emph{directional} FDR. We further analyze its statistical power (without imposing any assumption on the amplitude of the regression parameters or the noise level). 

Controlling FDR in regression model has been a long standing problem. It was just a couple years ago that~\cite{barber2015controlling} proposed the ingenious idea of knockoff filter. In a nutshell, this approach constructs a set of ``knockoff" variables that are  irrelevant to response (conditional on the original covariates) but whose structure mirrors that of the original covariates. The knockoff variables then behave as the controls for original covariates. This way, they bypass the need of constructing $p$-values and directly select a model with the desired FDR.  The focus of~\cite{barber2015controlling} was on linear regression model with $n>2p$. Later,~\cite{candes2018panning} extended the idea of knockoff filter to high-dimensional nonlinear models with random designs, but assumes that the joint distribution of covariates is known. Very recently,~\cite{fan2017rank,barber2018robust} studied robustness of model-X knockoff to errors in estimating the joint distribution of covariates. A salient feature of the knockoff approach is that for $n\ge 2p$, it controls FDR in finite sample setting without requiring any assumption on the covariates. However, the extension model-X knockoffs~\cite{candes2018panning} requires the knowledge of the joint distribution of covariates.  Moreover, the knockoff approach does not provide valid $p$-values for the hypotheses regarding the model parameters. By contrast, the FCD method that we present in this paper controls FDR as long as $s_0 = o(\sqrt{n}/(\log p)^2)$, without requiring the joint distribution of covariates.  Furthermore, it comes with the valid $p$-values for individual model parameters. However, the FDR control is proved for the asymptotic regime where $n\to  \infty$. \footnote{A finite sample analysis of our method is possible but requires a more involved analysis and is out of the scope of the present work.}

\subsection{Problem Formulation}
Suppose we have recorded $n$ i.i.d observational units $(y_1,x_1), \dotsc, (y_n,x_n)$, with $y_i\in \reals$ representing response variables and $x_i\in \reals^p$
indicating the vector of explanatory variables on each sample, also referred to as features. We assume the classical linear regression model where the observations obey the following relation:
\begin{align}
y_i = \<\theta_0,x_i\> + w_i\,,
\end{align}
Here, $\theta_0\in \reals^p$ is the unknown vector of coefficients. The symbol $\<\cdot,\cdot\>$ denotes the standard inner product. Let $y = (y_1, \dotsc, y_n)^\sT$ and let $X\in \reals^{n\times p}$ denote the feature matrix that have $x_1^\sT, \dotsc, x_n^\sT$ as rows. Then, writing the linear regression model in matrix form, we obtain
 \begin{align}
\label{eq:linearmodel}
y = X\theta_0 + w\,, 
\end{align}

We assume that conditional on the design $X$, the noise variables $w_i$ are independent with 
\begin{align}\label{eq:noise}
\E(w_i| X) = 0\,, \quad \E(w_i^2|X) = \sigma^2\,, \quad \E\big(|w_i|^{2+a}\big|X\big) \le C \sigma^{2+a}\,, 
\end{align}
for some constants $C>0$, $a>2$.

We let $S\subseteq \{1,\dots, p\}$ denote the set of truly relevant feature variables among the many that have been recorded. This set corresponds to the support of $\theta_0$, i.e., 
\begin{align}
S \equiv \supp(\theta_0) = \{1\le i\le p:\, \theta_{0,i}\neq 0\}\,.
\end{align}
We let $s_0 = |S|$ be the size of support or in other words the number of true positives.

In this paper, we propose a framework to select a set $\hS$ of the feature variables, while controlling the directional false discovery rate (FDR) for the selected variables. This criterion is intimately rated to type S errors (S stands for sign). Type S error (a.k.a type III error) occurs when we  say, with confidence, that a comparison goes one way while it goes the other way~\cite{gelman2000type}. For example, we claim that $\theta_1>\theta_2$, with confidence, while in fact $\theta_1<\theta_2$. In other words, we mistakenly make a claim on the sign (direction) of $\theta_1-\theta_2$.
Gelman et. al.~\cite{gelman2000type} argue that type S error is a more relevant notion in many applications.  Tukey also conveys a similar message in~\cite{tukey1991philosophy} by arguing that the effects of $A$ and $B$, for any $A$ and $B$, are always different  (in some decimal precision) and hence instead of questioning whether there is any difference in two effects, the valid question should be about the direction in which effect of $A$ differs from that of $B$.

Motivated by this, we formally define directional FDR, denoted by $\FDRdir$. For a selected set $\hS$ of the features along with estimates $\hsign_j \in \{-1, +1\}$ of the sign of $\theta_{0,j}$, we define
\begin{align}\label{FDRdir}
\FDRdir = \E[\FDPdir]\,, \quad \quad  \FDPdir = \frac{|\{j\in \hS:\, \hsign_j \neq \sign(\theta_{0,j})\}|}{\max(|\hS|,1)} \,,
\end{align}
where we adopt the convention $\sign(0) = 0$. In words, $\FDRdir$ is the expected fraction of false discoveries among the selected ones, where a false discovery is measured with respect to type S and type I errors. For example, if $\hsign_j = +1$, while $\theta_{0,j} = 0$ (type I error) or $\theta_{0,j} < 0$ (type S error), it is considered as a false discovery.

Recall that the classical FDR is defined as 
\begin{align}\label{FDR}
\FDR = \E\bigg[\frac{|\{j\in \hS:\, \theta_{0,j} \neq 0 \}|}{\max(|\hS|,1)} \bigg]\,,
\end{align}
that defines the false discoveries only with respect to type I error. Therefore, a comparison of definitions~\eqref{FDRdir} and~\eqref{FDR} reveals that  
\begin{align}
\FDRdir \ge \FDR\,,
\end{align}
for any selected set $\hS$. As a result, proving that a framework controls $\FDRdir$ automatically implies that it also controls FDR.

Likewise, we define the statistical power of a selected set $\hS$ as 
\begin{align}\label{power}
\power = \E\bigg[\frac{|\{j\in \hS:\, \hsign_j = \sign(\theta_{0,j})\}|}{\max(|S|,1)} \bigg]\,,
\end{align}
i.e., for a true discovery not only the corresponding variable should be in fact non-zero but we should also declare its sign correctly. 

The directional FDR has been also studied by~\cite{barber2016knockoff} and it is shown that the knockoff filter also controls this metric as well as the FDR.

\subsection{Our Contributions and Outline of the Paper}
Here, we provide a vignette of our contributions:
\begin{description}
\item {\bf Controlling directional FDR} In Section~\ref{sec:method}, we propose a method for selecting relevant variables using the debiasing approach. We use the acronym FCD to call this method (standing for ``FDR Control via Debiasing"). In Section~\ref{sec:theorem}, we show that for design matrices with subgaussian rows, under the standard condition $s_0 = o(\sqrt{n}/(\log p)^2)$, the FCD framework achieves exact directional FDR control. (See Theorem~\ref{thm:main} for a formal statement).

\item{\bf Characterizing the statistical power} In Section~\ref{sec:power}, we characterize the statistical power of the FCD method. In particular, for designs with subgaussian rows and a common precision matrix $\Omega\in\reals^{p\times p}$, we show that if the minimum nonzero coefficient, $\theta_{\min}$ satisfies $$\sqrt{n} \theta_{\min} - \sigma \sqrt{2(\max_{i\in [p]}\Omega_{ii})\log\left(\frac{2p}{qs_0}\right)} \to \infty\,,$$ then FCD achieves asymptotic power one. 

Recently,~\cite{fan2017rank} has studied the power of model-X knockoff filter, provided that $\theta_{\min} \sqrt{\frac{n}{\log p}}\to \infty$ and assuming a lower bound on the size of the model selected by the
knockoff procedure. Namely, if $|\hS| \ge cs_0$, for some constant $c\in (2(qs_0)^{-1},1)$. Under such assumptions, it is shown that the model-X knockoff approach achieves asymptotic power one. Other than being restrictive, these assumptions are hard to verify and a sufficient given condition is that the size of 
$\{j:\,|\theta_{0j}|\gg \sqrt{s_0 (\log p)/n}\}$ is at least $cs_0$, for some constant $c\in (2(qs_0)^{-1},1)$. This condition on the amplitude of nonzero coefficients is much stronger that the one we need for FCD to achieve power one.   

\item{\bf Numerical validation} We validate our approach on both synthetic and real data in Sections~\ref{sec:numerical} and \ref{sec:realdata} and compare its performance with the model-X knockoff. 
As  the simulations suggest, FCD method compares favorably to the model free knockoff in a wide range of setups. We also compare the statistical power of FCD with the theoretical characterization and show that they are in good agreement.
\end{description}

\noindent{\bf Techniques.} In our analysis of FDR, we use ideas from the debiasing approach~\cite{javanmard2013hypothesis,zhang2014confidence,javanmard2014confidence,van2014asymptotically,javanmard2013nearly} together with some results from~\cite{liu2013gaussian} regarding the order statistic of sum of Gaussian random variables (See Lemma 6.1, 6.2 therein.) It is worth mentioning that~\cite{liu2013gaussian} developed such results to use in the analysis of a method they proposed for Gaussian graphical model and its FDR. This context is very different from the problem studied in this paper and as expected the test statistics are also very different.  In our FCD approach, we construct the test statistics by debiasing the Lasso solution. These test statistics have a Gaussian part and a bias term. In applying the results from~\cite{liu2013gaussian}, we need to do a careful analysis of the bias term, and also the errors in noise level estimation. In addition, by a careful analysis of the test statistic and the data dependent threshold used in our procedure, we are able to analyze the statistical power of our approach. 

\subsection{Further Related Work}
There exists a copious theoretical literature developed on high-dimensional regression and the Lasso. Most existing studies have focused on prediction error ~\cite{GreenshteinRitov}, model selection properties~\cite{MeinshausenBuhlmann, zhao, Wainwright2009LASSO,candes2009near}, estimation consistency \cite{CandesTao, BickelEtAl}. For exact support recovery, it was known early on that, even in the classical setting of fixed $p$ and diverging $n$, support of Lasso will be different from $S$ (support of true signal) unless the columns of $X$, with index in $S$, are roughly orthogonal to the ones with index outside $S$~\cite{knight2000asymptotics}. This assumption is formalized under the so-called `irrepresentability condition'.
In a seminal work, Zhao and Yu~\cite{zhao} show that  this condition also allows exact support recovery in the high-dimensional setting ($p\gg n$). Independently,~\cite{MeinshausenBuhlmann} studied model selection problem for random Gaussian designs, with
applications to learning Gaussian graphical models. These papers consider the setting of $s_0 = O(n^c)$, for some $c<1$. Further, under a normalization of design such that its columns have norm at most $\sqrt{n}$, they require the minimum nonzero amplitude of the signal $\theta_{\min} = \min_{i\in S} \theta_{0,i}$ to satisfy $\theta_{min} > c\sqrt{s_0/n}$. Later,~\cite{Wainwright2009LASSO} improved these results for the case of random Gaussian designs and showed that for a broad range of covariance matrices, the Lasso can recover the support of a signal  for which $\theta_{\min} > c\sigma\sqrt{(\log p)/n}$. The model selection problem was also studied under the weaker, generalized irrepresentability condition, for the Gauss-Lasso estimator~\cite{javanmard2013model}.

As an alternative to irrepresentability condition,~\cite{lounici2008sup} proves the exact model selection under an incoherence assumption of $\max_{i\neq j} \hSigma_{ij} = O(1/s_0)$. This assumption is however stronger than irrepresentability condition~\cite{BuhlmannVanDeGeer}.

As discussed in the introduction, related to the model selection is the problem of hypothesis testing for high-dimensional regression. In~\cite{ZhangZhangSignificance,BuhlmannSignificance}, authors consider null hypotheses of form $H_{0,i}:\,\theta_{0,i} = 0$ and propose methods that achieve a given power $1-\beta$, if $|\theta_{0,i}| > c_\beta\sigma\sqrt{s_0(\log p)/n}$. Later,~\cite{javanmard2013hypothesis} proposed a method for random Gaussian designs, with known covariance, under the setting $s/p\to \eps$ and $n/p\to \delta$, for constants $\eps, \delta\in (0,1)$. The proposed method achieves a given power $1-\beta$, conditional on that $|\theta_{0,i}| > c_\beta\sigma/\sqrt{n}$. The debiasing approach~\cite{zhang2014confidence,javanmard2014confidence,van2014asymptotically} also has been proposed to test $H_{0,i}$ in the high-dimensional setting, with $s_0 = o(\sqrt{n}/(\log p))$. In~\cite{javanmard2014confidence}, it is shown that the debiasing based framework for testing $H_{0,i}$ achieves a given power $1-\beta$, if $\theta_{\min} > c_\beta \sigma\sqrt{(\log p)/n}$. Applicability of the debiasing approach is extended to the setting of $s_0 = o(n/(\log p)^2)$, for random Gaussian designs, using a `leave-one-out' technique~\cite{javanmard2015biasing}.

\subsection{Notations}
Here, we provide a summary of notations used throughout this paper. We use $[p] = \{1, \dotsc, p\}$ to refer to the first $p$ integers. For a vector $v$, we denote its coordinates by $v_i$ and let $v_S$ be the restriction of $v$ to indices in set $S$. Further, the term support of a vector indicates the nonzero coordinates of that vector, i.e., $\supp(v) = \{i\in [p]:\, v_i \neq 0\}$. We use $\id$ to denote the identity matrix and for clarity we might also make its dimension explicit as in $\id_{d\times d}$. For a matrix $A$, we denote its maximum and minimum singular values by $\sigma_{\max}(A)$ and $\sigma_{\min}(A)$, respectively. 
For a random vector $x$, we denote its subgaussian norm by $\|x\|_{\psi_2}$ defined as:
\[
\|X\|_{\psi_2} \equiv \sup_{q\ge 1} q^{-1/2} (E|X|^q)^{1/q}\,,
\] 
and for a random vector $X\in \reals^m$, its subgaussian norm is defined as $\|X\|_{\psi_2} = \sup_{u\in S^{m-1}} \|\<X,u\>\|_{\psi_2}$.
We use $\phi(z) = e^{-z^2/2}/\sqrt{2\pi}$ to refer to  the Gaussian density and  $\Phi(z) = \int_{-\infty}^z \phi(t) \de t$ to denote the Gaussian cumulative distribution. For two functions $f(n)$ and $g(n)$, with $g(n)\ge 0$, we write $f(n) =o(g(n))$ if $g(n)$ grows much faster than $f(n)$, i.e., $f/g \to 0$. We also write $f(n) = O(g(n))$,
if there exists a positive constant $C$ such that for all sufficiently large values of $n$, $|f(n)|\le C |g(n)|$.

\section{FCD Procedure: False Discovery Control via Debiasing}\label{sec:method}
In order to describe FCD framework, we first give an overview of debiasing approach. To this end, we focus on the Lasso estimator~\cite{Tibs96}, given by
\begin{align}
\label{eq:lassoestimator}
\htheta(y,X;\lambda) \equiv \arg\min_{\theta\in \reals^p} \bigg\{\frac{1}{2n} \|y-X\theta\|_2^2 + \lambda \|\theta\|_1 \bigg\}
\end{align} 
In case the optimization has more than one optimizer we select one of them arbitrarily. We will often drop the arguments $y, X$, as they are clear from the context.
There is a vast literature on the properties of the Lasso estimator in the high-dimensional regime ($n<p$), mainly through the lens of point estimation and prediction. A major quantity that plays a key role in the estimation error is the co-called \emph{Compatibility constant} of the design matrix $X$.  Let $\hSigma \equiv X^\sT X/n$ be the sample covariance matrix. In the high-dimensional setting, where $n<p$, $\hSigma$ is always singular, and this makes the estimation of $\theta_0$ challenging  since for the parameter family $\{\theta = \theta_0 + v\}$, with $v$ in the null-space of $\hSigma$, we have $X\theta = X\theta_0$ and hence we get the same response vector. A common assumption to cope with this problem is requiring $\hSigma$ to be nonsingular for a restricted set of directions.
\begin{definition}
For a symmetric matrix $\hSigma\in \reals^{p\times p}$ and a set $S\subseteq[p]$, the corresponding compatibility constant is defined as
\[
\phi^2(\hSigma,S) \equiv \min_{\theta\in \reals^p} \Big\{\frac{|S|\<\theta,\hSigma \theta\>}{\|\theta_S\|_1^2}:\; \theta\in \reals^p, \, \|\theta_{S^c}\|_1\le 3 \|\theta_{S}\|_1  \Big\}\,.
\] 
The matrix $\hSigma\in \reals^{p\times}$ is said to satisfy the \emph{compatibility condition} if $\phi(\hSigma, S)\ge \phi_0$.  
\end{definition}

Despite the great properties of Lasso in terms of point estimation and prediction, it is biased due to the $\ell_1$ penalty term. Indeed, bias is unavoidable in high-dimensional setting ($n<p$) as one needs to produce a point estimate, in $p$ dimension, from the observed data in lower dimension, $n$. Furthermore, characterizing the exact distribution of regularized estimator is  in general not tractable.  To deal with these challenges, the debiasing approach aims at first removing the bias of Lasso and producing an estimator that is amenable to distributional characterization.

\subsection{Debiasing Lasso}
A debiased estimator $\dth$ takes the general simple form of 
\begin{align}
\label{eq:debiasedest}
\dth = \hth + \frac{1}{n} M X^\sT (y - X\htheta)\,.
\end{align}
Here, $M$ is a `decorrelating' matrix. There are various proposals for constructing $M$; see e.g.~\cite{zhang2014confidence,javanmard2014confidence,van2014asymptotically}. In this paper we use the 
 construction introduced by \cite{javanmard2014confidence}. Here, we assume that the noise $w$ is Gaussian and then discuss the non-Gaussian case in Section~\ref{nongaussian}.
 
 To set the stage to describe construction of $M$, note that by plugging in for $y = X\theta_0+w$, we have
\begin{align}
\sqrt{n}(\dth - \theta_0) = \sqrt{n} (M\hSigma - \id) (\theta_0-\htheta) + \frac{1}{\sqrt{n}} MX^\sT w\,,
\end{align}
where $\hSigma \equiv (X^\sT X)/n$ is the empirical covariance of the feature vectors. The first term is the bias and is controlled by $|M\hSigma - \id|_\infty$, with $|\cdot|$ denoting the entrywise $\ell_\infty$ norm. The second term is the unbiased Gaussian noise whose covariance works out at $M\hSigma M^\sT$. The decorrelating matrix $M$ is constructed via a convex optimization that aims at reducing bias and variance of the coordinates of $\dth$ at the same time.

Construct $M = (m_1, m_2, \dots, m_p)^\sT \in \reals^{p\times p}$
by letting $m_i \in \reals^p$
be a solution to the following convex program
\begin{align}
\label{eq:Moptimize}
\begin{split}
\minimize \quad\quad &m^\sT \hSigma m\,,\\
\subjectto \quad \quad &\|\hSigma m - e_i\|_\infty \leq \mu\,,
\end{split}
\end{align}
with $e_i\in \reals^p$ being the $i$'th standard unit vector. If any of the above problems 
is not feasible, we let $M = \id_{p\times p}$. Note that $M$ is constructed solely based on $X$. The choice of running parameter $\mu$ will be discussed in the sequel.

The following proposition proved in \cite{javanmard2014confidence}
shows that the error of the debiased estimator $\dth$ can be decomposed as the sum of two `bias' and
`noise' terms. In addition, a high probability bound is established on the bias term $\|\Delta\|_\infty$, which leverages on the properties of the optimization~\eqref{eq:Moptimize} and the estimation error of the Lasso estimator. Note that the compatibility condition for the design matrix $X$ is required for Lasso to achieve optimal estimation rate in high dimension~\cite{buhlmann2011statistics,BuhlmannVanDeGeer}. In~\cite{javanmard2014confidence}, there is also a version of the following proposition stated for deterministic results, with the compatibility constant $\phi_0$ explicit in the bound (see Theorem 2.3 therein.) The next proposition concerns the setting of random designs, which per se implies the compatibility condition. Indeed, by employing a reduction principle established by~\cite{rudelson2011reconstruction}, if the population covariance $\Sigma$ has minimum singular value $c_{\min} >0$ and provided a large enough sample size, namely $n \ge C s_0 \log(p/s_0)$, the sample covariance $\hSigma$ satisfies the compatibility condition with constant $\phi_0 = \sqrt{c_{\min}}/2$, with high probability.

\begin{propo}
\label{propo:biasvardecompose}
Consider the linear model \eqref{eq:linearmodel}, with gaussian noise, $w\sim \normal(0,\sigma^2 \id_{n\times n})$,  and let $\dth$ be the debiased estimator
given by Eq. \eqref{eq:debiasedest}, with $\mu = a\sqrt{(\log p)/n}$. Then, we have the following decomposition:
\begin{align}
\label{eq:biasvariancedec}
\sqrt{n}(\dth - \theta_0) = Z + \Delta,\quad
Z | X \sim \normal(0, \sigma^2 M \hSigma M^\sT), \quad \Delta = \sqrt{n}(M\hSigma - \id)(\theta_0-\hth).
\end{align}
Consider random design matrices with i.i.d rows and let $\Sigma = \E(x_1x_1^\sT)$ be the population level covariance. Suppose that $\sigma_{\min}(\Sigma) \ge c_{\min} > 0$ and $\sigma_{\max}(\Sigma)<c_{\max}$, for some constants $c_{\min}$, $c_{\max}$ and $\max_{i\in [p]} \Sigma_{ii}\le 1$. Further, assume that $X\Sigma^{-1/2}$ has independent subgaussian rows with $\|\Sigma^{-1/2} x_1\|_{\psi_2} \le \kappa$. Then, choosing $\lambda = c\sigma \sqrt{(\log p)/n}$, there exists constant $C= C(a,\kappa)$, such that for $n\ge Cs_0\log(p/s_0)$, we have
\begin{align}
\prob\left\{\left\|\Delta\right\|_\infty\geq \left(\frac{16ac_0\sigma}{c_{\min}}\right)\frac{s_0\log p}{\sqrt{n}}\right\} \le 4e^{-c_1 n} + 4p^{-c_2}\,,
\end{align}
where $c_1$ and $c_2$ are constants depending on $\kappa, a, c_0, c_{\min}, c_{\max}$.
\end{propo}

The next lemma controls the variance of the noise coordinates $Z_i$ in terms of the diagonal entries of the precision matrix.  

\begin{lemma}[~\cite{javanmard2014confidence}]\label{lem:app}
Let $\Omega\equiv \Sigma^{-1}$ be the precision matrix. Under the assumption of Proposition~\ref{propo:biasvardecompose}, the following holds true for any fixed sequence of integers $i = i(n)$:
\begin{align}
\prob\left(m_i^\sT \hSigma m_i - \Omega_{i,i}\ge \epsilon \right) \le 2 e^{-(n/6) (\epsilon/e\kappa')^2} + 2p^{-c}\,,
\end{align}
for $\kappa'\equiv 2\kappa^2 c_{\min}^{-1}$ and a constant $c = c(a)>0$.
\end{lemma}

\subsection{Extension to Non-Gaussian Noise}\label{nongaussian}
In the decomposition~\eqref{propo:biasvardecompose}, we have $Z = MX^\sT W/\sqrt{n}$ and given that $W\sim\normal(0,\Sigma)$, we have $Z|X\sim\normal(0,\sigma^2 M\hSigma M^\sT)$. In~\cite{javanmard2014confidence}, it is shown that by a slight modification of optimization~\eqref{eq:Moptimize}, $Z$ admits the same conditional distribution even for non-Gaussian noise. For the reader's convenience and to be self-contained we briefly explain it here.

Note that for any fixed $i\in[p]$, we have 
\[
Z_i= \frac{1}{\sqrt{n}} \sum_{j=1}^n \xi_j\,, \quad \text{ with }\quad \xi_j \equiv \frac{m_i^\sT x_j w_j}{\sigma (m_i^\sT \hSigma m_i)^{1/2}}\,. 
\]
Conditional on $X$, the terms $\xi_j$ are zero mean and independent. Moreover, $\sum_{j=1}^n \E(\xi_j^2|X) = n$. Therefore, if the Lindeberg's condition holds, that is to say for every $\eps>0$, almost surely
\[
\lim_{n\to \infty} \frac{1}{n}\sum_{j=1}^n \E(\xi_j^2\ind(|\xi_j| > \eps \sqrt{n})|X) = 0\,,
\]
then $\sum_{j=1}^n \xi_j/\sqrt{n}|X {\dist}\normal(0,1)$. The construction of $M$ can slightly be modified to ensure the Lindeberg's condition, namely optimization problem~\eqref{eq:Moptimize} should be modified as follows:
\begin{align}
\label{eq:Moptimizemodified}
\begin{split}
\minimize \quad\quad &m^\sT \hSigma m\,,\\
\subjectto \quad \quad &\|\hSigma m - e_i\|_\infty \leq \mu\,,\\
&\|Xm\|_\infty \le n^\beta\quad \text{for arbitrary fixed } 0<\beta<1/2- a^{-1}\,, 
\end{split}
\end{align}
where we recall the parameter $a$ from Eq.~\eqref{eq:noise}. The following lemma, which is from~\cite{javanmard2014confidence}, shows that by this modification, the marginals of $Z_j$ are asymptotically normal. 

\begin{lemma}[\cite{javanmard2014confidence}, Theorem 4.1]
Under conditions~\eqref{eq:noise} on the noise term, and using optimization~\eqref{eq:Moptimize} to construct $M$, we have that $Z_i|X\dist\normal(0,1)$.
\end{lemma}
The above result can be easily generalized to fixed-dimensional marginals of $Z$, by using the fact that a vector has a multivariate normal distribution
if every linear combination of its coordinates is normally distributed. 

With this overview of debiasing approach we are ready to explain the FCD procedure.

\subsection{FCD Procedure}
\label{subsec:FCD}
\subsubsection{Construction of Test Statistics}
In order to construct the test statistics, we first need to propose a consistent estimate of noise variance, $\sigma^2$. There are already several proposals for this in the literature. See e.g.,~\cite{SCAD01,sis08,SBvdG10,mcp10,SZ-scaledLasso,
BelloniChern,FGH12,reid2016study}.
To be concrete, we use the scaled Lasso~\cite{SZ-scaledLasso} given by
\begin{equation}\label{scLasso}
\{\htheta,\hsigma\} \equiv \arg\min_{\theta\in \reals^p, \sigma>0}  \bigg\{\frac{1}{2\sigma n} \|y-X\theta\|_2^2 + \frac{\sigma}{2} + \bar{\lambda} \|\theta\|_1 \bigg\}
\end{equation} 
We state the following lemma that shows $\hsigma$ is a consistent estimate of $\sigma$. We refer to~\cite[Lemma 3.3]{javanmard2014confidence} or~\cite[Theorem 1]{SZ-scaledLasso} for its proof.

\begin{lemma}
\label{lemma:sigmahatoversigma}
Consider a sequence of design matrices $X\in \reals^{n\times p}$, with dimensions $n\to \infty$, $p= p (n)\to \infty$. For each $n$, let $\Sigma\in \reals^{p\times p}$ such that $\sigma_{\min}(\Sigma) \ge c_{\min} > 0$ and $\sigma_{\max}(\Sigma)<c_{\max}<\infty$, for some constants $c_{\min}$, $c_{\max}$ and $\max_{i\in [p]} \Sigma_{ii}\le 1$. Further, assume that $X\Sigma^{-1/2}$ has independent subgaussian rows, with zero mean and  subgaussian norm $\|\Sigma^{-1/2} x_1\|_{\psi_2} \le \kappa$. 
Let $\hsigma$ be the scaled Lasso estimate of the noise level, defined by~\eqref{scLasso}, with $\bar{\lambda} = 2\sqrt{(2\log p)/n}$.
Then, assuming $s_0 = o(n/(\log p))$, the estimator $\hsigma$ satisfies the following relation:
 \begin{eqnarray*}
 \lim_{n\to\infty} \underset{\theta_0\in \reals^p, \|\theta_0\|_0\le s_0}{\sup} \prob\bigg(\Big|\frac{\hsigma}{\sigma} - 1 \Big|\ge \eps \bigg) = 0\,.
 \end{eqnarray*}
 Here, $\prob$ is w.r.t the randomness of the noise $w$ and the design $X$.
\end{lemma}

Define $\Lambda = M\hSigma M^\sT$. For $i\in [p]$, we define test statistic $T_i$ as follows:
\begin{eqnarray}\label{test-statistic}
T_i \equiv \frac{\sqrt{n} \dth_i}{\hsigma \sqrt{\Lambda_{ii}}}\,.
\end{eqnarray}

For a given threshold level $t\ge 0$, we reject $H_{0,i}$ if $|T_i|\ge t$
and we return sign of $T_i$ as the estimate of sign of $\theta_{0,i}$. 
We also let $R(t)=\sum_{i=1}^p \ind(|T_i|\ge t)$ be the total set of rejections at threshold $t$.
Next, we discuss how to choose a data dependent threshold $t$ to ensure
that directional FDR and FDP are controlled at a pre-assigned level $q\in [0,1]$.

\subsubsection{A Data Dependent Threshold for the Test Statistics}\label{subsec:fdrprocedure}

\begin{itemize}
\item Step 1: For the pre-assigned level $q\in [0,1]$, let $t_p = (2\log p - 2\log \log p)^{1/2}$ and calculate
\begin{align}\label{t0-1}
t_0 = \inf \bigg\{0\le t\le t_p:\, \frac{2p(1-\Phi(t))}{R(t) \vee 1} \leq q \bigg\}\,.
\end{align}
 If~\eqref{t0-1} does not exist then set $t_0 = \sqrt{2\log p}$.

\item Step 2: For $i\in [p]$, reject $H_{0,i}$ if $|T_i|\ge t_0$.

\item Step 3: We return $\hsign_i = \sign(T_i)$ as the estimate of $\sign(\theta_{0,i})$.
\end{itemize}


\section{Main Results}\label{sec:theorem}
\subsection{Control of Directional False Discovery Rate}
Suppose that the design matrix $X$ has i.i.d rows with $\Sigma  = \E(x_1x_1^\sT)$ being the population covariance. Let $\Omega\equiv \Sigma^{-1}$ be the precision matrix and recall the definition $\Lambda \equiv M\hSigma M^\sT$, where $M$ is the decorrelating matrix used in construction of the debiased estimator.

We also define the normalized matrices $\Omega^0$ and $\Lambda^0$ as 
\begin{align}
\Omega^0_{ij} = \frac{\Omega_{ij}}{\sqrt{\Omega_{ii} \Omega_{jj}}}\,, \quad 
\Lambda^0_{ij} = \frac{\Lambda_{ij}}{\sqrt{\Lambda_{ii} \Lambda_{jj}}}\,.
\end{align}
For a given constant $\gamma > 0$,
define
\begin{align}
\label{eq:Lambda(gamma)}
&\Gamma(\gamma,c_0) \equiv \Big\{(i,j): 1\le i, j\le p, |\Omega^0_{ij}|\ge c_0(\log p)^{-2-\gamma} \Big\}\,,
\end{align}
for some constant $c>0$. The following theorem states a guarantee on the directional false discovery rate of the 
FCD procedure introduced in the previous section. 

\begin{thm}\label{thm:main}
Consider random design matrices with i.i.d rows and let $\Sigma = \E(x_1x_1^\sT)$ be the population level covariance. Suppose that $\sigma_{\min}(\Sigma) \ge c_{\min} > 0$ and $\sigma_{\max}(\Sigma)<c_{\max}$, for some constants $c_{\min}$, $c_{\max}$ and $\max_{i\in [p]} \Sigma_{ii}\le 1$. In addition, assume that $X\Sigma^{-1/2}$ has independent subgaussian rows with $\|\Sigma^{-1/2} x_1\|_{\psi_2} = \kappa$. Also assume that:
\begin{itemize}
\item[$(i)$] $s_0 = o(\sqrt{n}/(\log p)^2)$.
\item[$(ii)$] There exist positive constants $c_0$, $\gamma$, such that $|\Gamma(\gamma,c_0)| = o(p^{1+\rho})$, for some constant $\rho\in [0,1)$.
\item[$(iii)$] We have $|\{(i,j): |\Omega_{ij}^0| > (1-\rho)/(1+\rho)\}| = O(p)$.
\end{itemize}
Then, for FCD procedure we get
\begin{align}\label{FDR_dir_limit}
\underset{(n,p) \to \infty}{\lim\sup}\, \FDRdir \le q.
\end{align}
Further, for any $\eps > 0$,
\begin{align}\label{eq:FDP}
\lim_{(n,p)\to \infty} \prob\Big(\FDPdir \le q+\eps \Big) = 1\,.
\end{align}
\end{thm}

\begin{remark}
While directional FDR is the \emph{expected} directional false discovery proportion ($\FDPdir$), it is idealized for a variable selection procedure to control $\FDPdir$ in any given realization. In general, controlling $\FDRdir$ does not control the variations of $\FDPdir$. As noted by~\cite{owen2005variance}, the
variance of FDP can be large if the test statistics are correlated, which is the case here. Let us emphasize that by Eq.~\eqref{eq:FDP}, our FCD controls $\FDPdir$, with high probability.   
\end{remark}

\bigskip

\noindent{\bf Examples.} Here, we provide several examples of the precision matrices that satisfy conditions $(ii)$-$(iii)$ of Theorem~\ref{thm:main} to demonstrate its applicability.
\begin{description}
\item {\emph{Example 1:}} Our first example is the circulant covariance matrices, where $\Sigma_{ij} = \eta^{|i-j|}$, for some constant $\eta\in (0,1)$.
It is simple to see that the inverse of such matrices has at most three nonzero coordinates per row. Therefore, the conditions will be satisfied by choosing $\rho =1$, and $c>0, \gamma <\infty$, arbitrary.
\item{\emph{Example 2:}} Suppose that $\Sigma$ is block diagonal with size of blocks to bounded (as $p\to \infty$). Then, the precision matrix will also have a block diagonal structure with blocks of bounded size. It is easy to check conditions, with choosing $\rho = 1 $ and $c>0, \gamma <\infty$, arbitrary. 
\item{\emph{Example 3:}} Consider the equi-correlated features, where $\Sigma = (1-r) \id + r {\mathbf 1}{\mathbf 1}^\sT$, for some constant $r\in (0,1)$, where ${\mathbf 1}\in \reals^p$ denotes the all-one vector. Then, we have $\Omega = (a-b)\id+ b{\mathbf 1}{\mathbf 1}^\sT$, with 
\begin{align}
a = \frac{(p-2)r+1}{(p-2)r-(p-1)r^2+1}\,, \quad b = \frac{-r}{(p-2)r - (p-1)r^2+1}\,. 
\end{align}
Note that $|b| = O(1/p)$. Therefore, the conditions hold for arbitrary constants $c>0$, $0 <\rho<1$.    
\end{description}

Finally, consider two matrices $\Omega^{(1)}$ and $\Omega^{(2)}$, with same diagonal entries $\Omega^{(1)}_{ii} = \Omega^{(2)}_{ii}$, for $i\in [p]$, such that $\Omega^{(1)}$ dominates $\Omega^{(2)}$ on off-diagonal entries, i.e., $\Omega^{(1)}_{ij} \geq \Omega^{(2)}_{ij}$, for $i\neq j\in [p]$. Then it is easy to see that if $\Omega^{(1)}$ satisfies Conditions $(i)$-$(ii)$, so does $\Omega^{(2)}$.
\subsection{Power Analysis}\label{sec:power}
Recall that $S_0\equiv \supp(\theta_0)$ is the set of indices of the truly significant features. Let $\hS$ denote the set of significant parameters returned by our FCD procedure, namely
\begin{align}
\hS = \{1\le j\le p:\, |T_j|\ge t_0\}\,. 
\end{align}
The power of a selected model $\hS$ is defined as
\begin{align}\label{powerShat}
\power(\hS) = \E\bigg[\frac{|\{j\in \hS:\, \hsign_j = \sign(\theta_{0,j})\}|}{\max(|S|,1)} \bigg]\,.
\end{align}
We are now ready to characterize the statistical power of the FCD procedure for the
linear model~\eqref{eq:linearmodel}. 

\begin{thm}\label{thm:power}
Consider a sequence of random design matrices $X\in \reals^{n\times p}$, with dimension $n\to\infty$, $p = p(n)\to\infty$ and $\Sigma = \E(x_1x_1^\sT)\in \reals^{p\times p}$. Suppose that $\sigma_{\min}(\Sigma) \ge c_{\min} > 0$ and $\sigma_{\max}(\Sigma)<c_{max}$, for some constants $c_{\min}$, $c_{\max}$ and $\max_{i\in [p]} \Sigma_{ii}\le 1$. Further, assume that $X\Sigma^{-1/2}$ has independent subgaussian rows with $\|\Sigma^{-1/2} x_1\|_{\psi_2} = \kappa$.
Suppose that $s_0 = o(\sqrt{n}/(\log p)^2)$ and for $i\in S=\supp(\theta_0)$, we have $|\theta_{0,i}| > ({\sigma}/\sqrt{n}) \sqrt{2\Omega_{ii}\log(p/s_0)}$. Then, the following holds true:
\begin{align}
&\underset{n\to\infty}{\lim\inf} \frac{\power(\hS)}{1-\beta(\theta_0,n)} \ge 1\\
&1-\beta(\theta_0,n) = \frac{1}{s_0} \sum_{i\in S} F\left(\frac{qs_0}{p}, \frac{\sqrt{n}|\theta_{0,i}|}{\sigma\sqrt{\Omega_{ii}}}\right)\,,
\end{align}
where, for $\alpha\in [0,1]$ and $u\in \reals_+$, the function $F(\alpha,u)$ is defined as follows:
\begin{align}\label{def:G}
F(\alpha, u) \equiv 1 - \Phi(\Phi^{-1}(1-{\alpha}/{2})-u) \,.
\end{align}
\end{thm}
We refer to Section~\ref{proof:thm-power} for the proof of Theorem~\ref{thm:power}.

\begin{coro}\label{coro:theta-min-power}
It is easy to see that for any fixed $\alpha\in [0,1]$, function $u\mapsto F(\alpha,u)$ is monotone increasing. Therefore, as a result of Theorem~\ref{thm:power}, we have
\begin{align}
\underset{n\to\infty}{\lim\inf} \frac{\power(\hS)}{F\left(\dfrac{qs_0}{p},\dfrac{\sqrt{n}\theta_{\min}}{\sigma\sqrt{\Omega_{ii}}}\right)}\ge 1\,.
\end{align} 
\end{coro}

\begin{coro}\label{coro:unit-power}
Under the assumptions of Theorem~\ref{thm:power}, if 
\begin{align*}
\sqrt{n} \theta_{\min} - \sigma \sqrt{2\max_{i\in [p]}(\Omega_{ii})\log(2p/(qs_0))} \to \infty,
\end{align*}
then $\power(\hS) \to 1$, as $n\to \infty$. 
\end{coro}
Proof of Corollary~\ref{coro:unit-power} is given in Appendix~\ref{proof:unit-power}.


\section{Improved Results for Gaussian Designs}
In~\cite{javanmard2015biasing}, the authors improved upon Proposition~\ref{propo:biasvardecompose} for Gaussian designs by providing a sharper bound for $\|\Delta\|_\infty$ using a `leave-one-out' technique. Specifically, for  Gaussian designs with known population covariance, it is shown that $\|\Delta\|_\infty = o_p(\sqrt{\frac{s_0}{n}} \log p)$. The same bound holds when the population covariance is unknown but can be estimated sufficiently well. e.g., if the inverse covariance  is sufficiently sparse. In this section, we aim at employing this result to relax the sparsity assumption (Condition (i)) in Theorem~\ref{thm:main}.    
\subsection{Known Covariance}
Consider linear model~\eqref{eq:linearmodel} where the design $X$ has independent Gaussian rows, with zero mean and covariance $\Sigma$. Also, denote by $\Omega\equiv \Sigma^{-1}$ be the inverse population covariance, a.k.a precision matrix. Here, we assume that $\Sigma$ is known and consider the test statistic $T_i$, given by~\eqref{test-statistic} where $\dth$ is the debiased estimator with $M = \Omega$.

For an integer $1\le k\le p$, define $\tau(\Omega,k)$ as follows:\footnote{In~\cite{javanmard2015biasing}, the authors use the notation $\rho(\Omega,k)$ to refer to the same quantity. We avoid that notation as we have used the symbol $\rho$ in Condition (iii) in Theorem~\ref{thm:main2}.}
\[
\tau(\Sigma,k)\equiv \max_{A\subseteq[p],|A|\le k} \|(\Sigma_{A,A})^{-1}\|_\infty\,,
\]
where $\|\cdot\|_\infty$ denotes the $\ell_\infty$ operator norm (maximum $\ell_1$ norm of the rows). As proved in~\cite{javanmard2015biasing}, we have the following bound in place:
\[
\tau(\Sigma,k) \le \min \Big\{\|\Omega\|_\infty, \sqrt{k} \sigma_{\max}(\Omega) \Big\}\,.
\] 
The next theorem is analogues to Theorem~\ref{thm:main} for Gaussian designs, under a weaker assumption on the sparsity level $s_0$.
\begin{thm}\label{thm:main2} (Known covariance).
Consider a sequence of Gaussian random design matrices $X\in \reals^{n\times p}$, with dimension $n\to\infty$, $p = p(n)\to\infty$. Suppose that $X$ has  i.i.d rows with zero mean and $\Sigma = \E(x_1x_1^\sT)$ be the population covariance. Suppose that $\sigma_{\min}(\Sigma) \ge c_{\min} > 0$ and $\sigma_{\max}(\Sigma)<c_{\max}$, for some constants $c_{\min}$, $c_{\max}$ and $\max_{i\in [p]} \Sigma_{ii}\le 1$. Further, assume that:
\begin{itemize}
\item[$(i)$] $s_0 = o({n}/(\log p)^4)$.
\item[$(ii)$] Let $C_0 = (32 c_{\max}/c_{\min})+1$. We have $\tau(\Sigma, C_0 s_0)\le \tau_0$, for some constant $\tau_0>0$.
\item[$(iii)$] There exist positive constants $c_0$, $\gamma$, such that $|\Gamma(\gamma,c_0)| = o(p^{1+\rho})$, for some constant $\rho\in [0,1)$.
\item[$(iv)$] We have $|\{(i,j): |\Omega_{ij}^0| > (1-\rho)/(1+\rho)\}| = O(p)$.
\end{itemize}
Then, for FCD procedure we get
\begin{align}\label{FDR_dir_limit_known}
\underset{(n,p) \to \infty}{\lim\sup}\, \FDRdir \le q.
\end{align}
Further, for any $\eps > 0$,
\begin{align}\label{eq:FDP_known}
\lim_{(n,p)\to \infty} \prob\Big(\FDPdir \le q+\eps \Big) = 1\,.
\end{align}
\end{thm}
The proof of Theorem~\ref{thm:main2} proceeds along the same lines as proof of Theorem~\ref{thm:main} and uses the result of~\cite[Theorem 3.8]{javanmard2015biasing}. We refer to section~\ref{proof:thm-main2} for its proof. 
\subsection{Unknown Covariance}
For the case of unknown covariance, we follow the construction of the decorrelating matrix $M$ proposed in~\cite{van2014asymptotically}. This construction is based on node-wise Lasso on matrix $X$. Formally, for $i\in [p]$, let $\tx_i$ be the $i$-th column of $X$ and represent it via sparse regression against all other columns:
\[
\hgamma_i(\tilde{\lambda}) = \underset{\gamma\in \reals^p}{\arg\min}\Big\{\frac{1}{2n} \|\tx_i - X_{\sim i} \gamma\|_2^2 + \tilde{\lambda} \|\gamma\|_1\Big\}\,,
\]     
where $X_{\sim i}$ is the submatrix obtained by removing the $i$-th column. Let 
\begin{align*}
\widehat{C} = \begin{bmatrix}
1& - \hat{\gamma}_{1,2}& \cdots & -\hat{\gamma}_{1,p}\\
- \hat{\gamma}_{2,1}& 1&\cdots & - \hat{\gamma}_{2,p}\\
\vdots & \vdots&\ddots& \vdots\\
- \hat{\gamma}_{p,1} & - \hat{\gamma}_{p,2} & \cdots & 1
\end{bmatrix}\,.
\end{align*}

Also define 
\begin{align}
\widehat{T}^2 = \diag(\htau_1^2, \dotsc, \htau_p^2)\,, \quad \quad \htau_i^2 = \frac{1}{n} (\tx_i - X_{\sim i} \hgamma_i)^\sT \tx_i\,.
\end{align}
The decorating matrix $M$ is then defined as 
\begin{align}\label{eq:M}
M \equiv \widehat{T}^{-2}\widehat{C}\,.
\end{align}
We consider the FDC procedure, where the test statistic $T_i$ is given by~\eqref{test-statistic} and $\dth$ is the debiased estimator with the decorrelating matrix $M$~\eqref{eq:M}. 

Define the sparsity level $s_\Omega$ for the precision matrix $\Omega$ as:
\[
s_\Omega \equiv \max_{i\in[p]} \big|\{j\neq i, \, \Omega_{i,j}\neq 0\} \big|\,.
\]
In words, $s_\Omega$ is the maximum sparsity of the rows of $\Omega$.

For the case of Gaussian designs with unknown covariance, we prove that the directional FDR of the FCD procedure is controlled under a weaker assumption on the sparsity of the parameters $s_0$, provided $s_\Omega$ is small enough.
\begin{thm}\label{thm:main3}(Unknown covariance).
Consider a sequence of Gaussian random design matrices $X\in \reals^{n\times p}$, with dimension $n\to\infty$, $p = p(n)\to\infty$ and $X$ has  i.i.d rows with zero mean and covariance $\Sigma$. Assume that $\sigma_{\min}(\Sigma) \ge c_{\min} > 0$ and $\sigma_{\max}(\Sigma)<c_{\max}$, for some constants $c_{\min}$, $c_{\max}$ and $\max_{i\in [p]} \Sigma_{ii}\le 1$. Further, suppose that 
\begin{itemize}
\item[$(i)$] $s_0 = o({n}/(\log p)^4)$ and $\min(s_0,s_\Omega) = o(\sqrt{n}/(\log p)^2)$. 
\end{itemize}
and Conditions $(ii)$, $(iii)$, $(iv)$ in Theorem~\ref{thm:main2} hold for $\Sigma$. 
Then, for FCD procedure we get
\begin{align}\label{FDR_dir_limit_unknown}
\underset{(n,p) \to \infty}{\lim\sup}\, \FDRdir \le q.
\end{align}
Further, for any $\eps > 0$,
\begin{align}\label{eq:FDP_unknown}
\lim_{(n,p)\to \infty} \prob\Big(\FDPdir \le q+\eps \Big) = 1\,.
\end{align}
\end{thm}
The proof of Theorem~\ref{thm:main3} proceeds along the same lines as proof of Theorem~\ref{thm:main} and uses the result of~\cite[Theorem 3.13]{javanmard2015biasing}. We refer to section~\ref{proof:thm-main3} for its proof. 

\section{Numerical Experiments}\label{sec:numerical}
We consider linear model~\eqref{eq:linearmodel} where the design matrix $X$ is generated by drawing its rows independently from $\normal(0,\Sigma)$. The covariance $\Sigma\in\reals^{p\times p}$ has a circulant structure with $\Sigma_{ij} = \eta^{|i-j|}$, for some constant $\eta\in(0,1)$. We then normalize the columns of $X$ to have unit norm. We generate the vector of coefficients $\theta_0\in \reals^p$ by choosing a subset of indices $S\subseteq [p]$ at random, of size $s_0$ and setting $\theta_{0,i}$ from $\{\pm A\}$ uniformly at random and $\theta_{0,i}  = 0$, for $i\notin S_0$.  The noise term $W$ is drawn from $\normal(0,\id_{n\times n})$. 

We perform three sets of simulations to compare the performance of FCD procedure with model free knockoff and to examine the effects of sparsity
level, signal magnitude, and feature correlation. We also compare the empirical power of FCD with the analytical lower bound provided in Corollary~\ref{coro:theta-min-power}. In all simulations, we set the target level FDR to $q = 0.1$.

For FCD procedure, we use the implementation of the debiased method provided by \url{http://web.stanford.edu/ montanar/sslasso/}, to construct the debiased estimator. For model free knockoff, we use the implantation provided by \url{http://web.stanford.edu/group/candes/knockoffs/}.

\bigskip

\begin{figure}[]

\begin{tabular}{lll}
\includegraphics[width = 4.8cm]{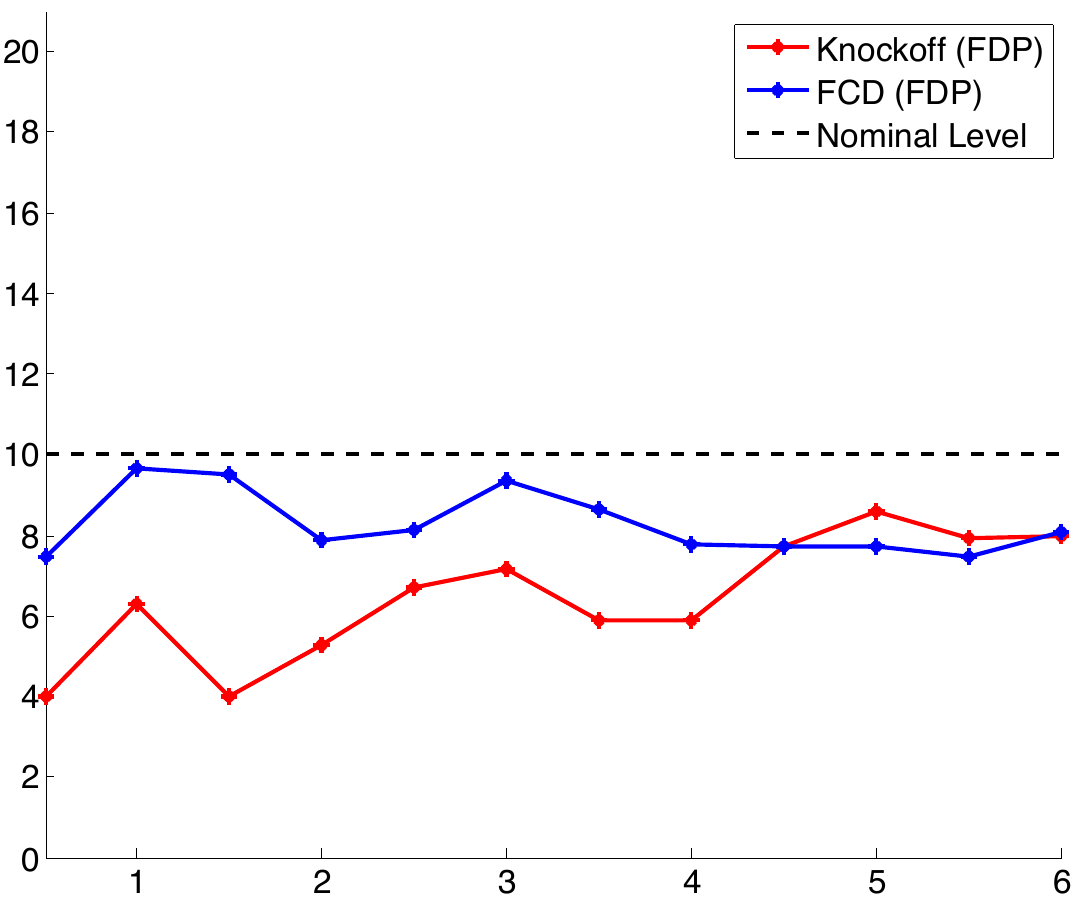}&\phantom{AAAAA}&
\includegraphics[width=4.8cm]{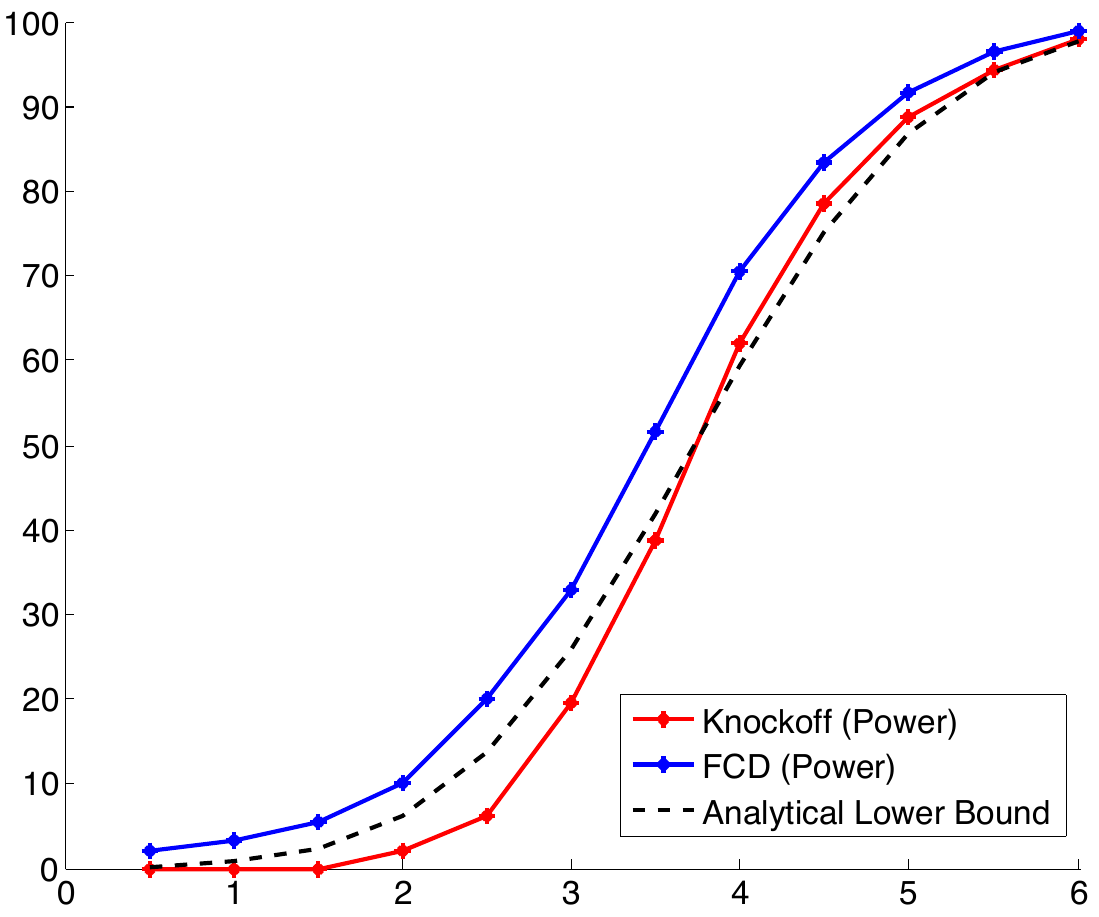}
\end{tabular}
\put(-290,-62){{\scriptsize Amplitude $(A)$}}
\put(-95,-62){{\scriptsize Amplitude $(A)$}}
\put(-345,-10){\rotatebox{90}{{\scriptsize $\FDRdir$}}}
\put(-150,-25){\rotatebox{90}{{\scriptsize statistical power}}}

\caption{Testing FCD and model free knockoff methods with varying the coefficients amplitude $A$. Here, $n = 2000$, $p = 3000$, $k = 100$, $\eta = 0.1$. The target level is $q = 10\%$. $\FDRdir$ and power are computed by averaging over $100$ realizations of noise and coefficient vectors.}\label{fig:amp}
\end{figure}

\noindent{{\bf Effect of Signal Amplitude:}}
We choose $n = 2000$, $p = 3000$, $k = 100$, $\eta = 0.1$ and vary the signal amplitude in the set $A\in \{0.5, 1, 1.5, \dotsc, 5.5, 6\}$. 
For the FCD procedure and the model free knockoff, we compute the directional FDR and power by averaging across $100$ realizations of noise and the generation of coefficient vector $\theta_0$.
 The results are plotted in Figure~\ref{fig:amp}. As we observe, both methods control $\FDRdir$ under the target level $q = 0.1$.
 As expected, the power of both procedures increases as the signal amplitude increases, with FCD procedure having larger power than the
 knockoff method over the entire range of signal amplitudes.
 The FCD procedure turn out to be more powerful than knockoff procedure. 
 
 We also plot the analytical lower bound on the power of FCD procedure, given in Corollary~\ref{coro:theta-min-power}. As we see the lower bound is quite close to the actual empirical power of FCD procedure in the setup tested.
 
\bigskip

\begin{figure}[]

\begin{tabular}{lll}
\includegraphics[width = 4.8cm]{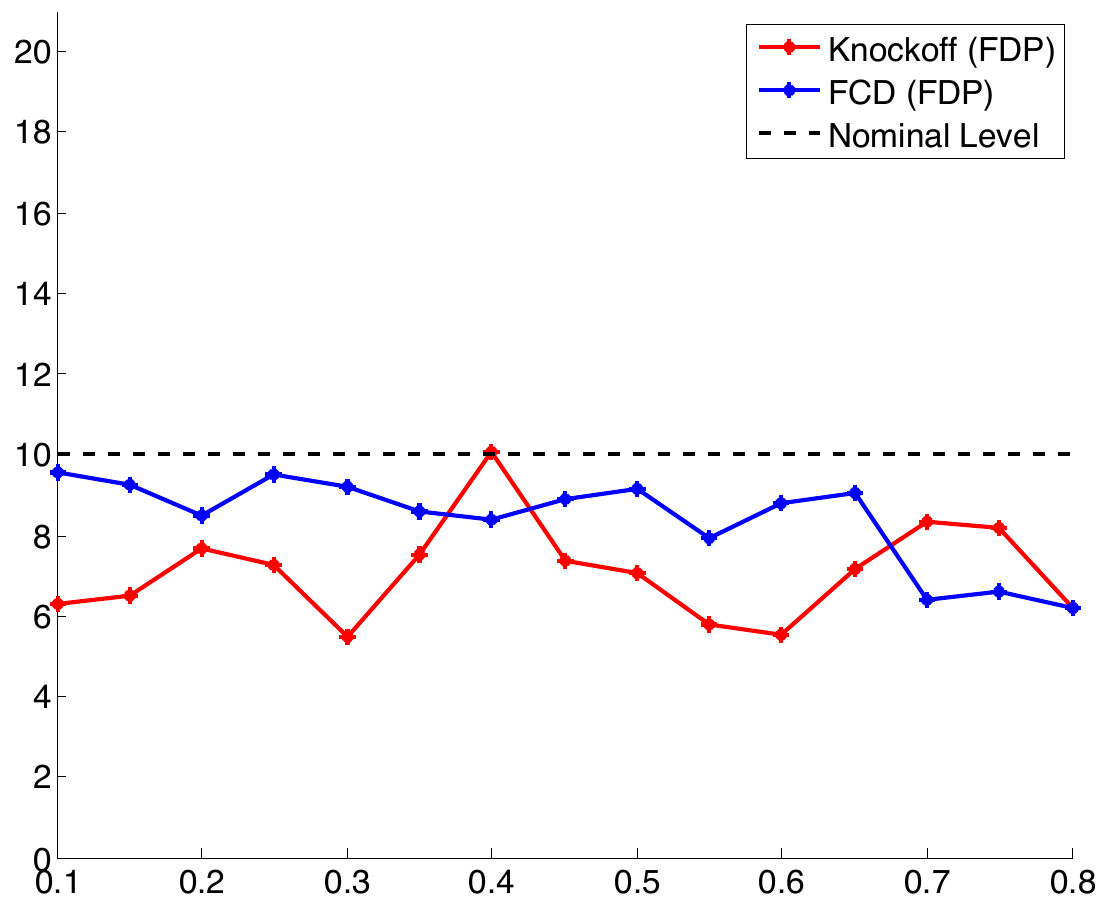}&\phantom{AAAAA}&
\includegraphics[width=4.8cm]{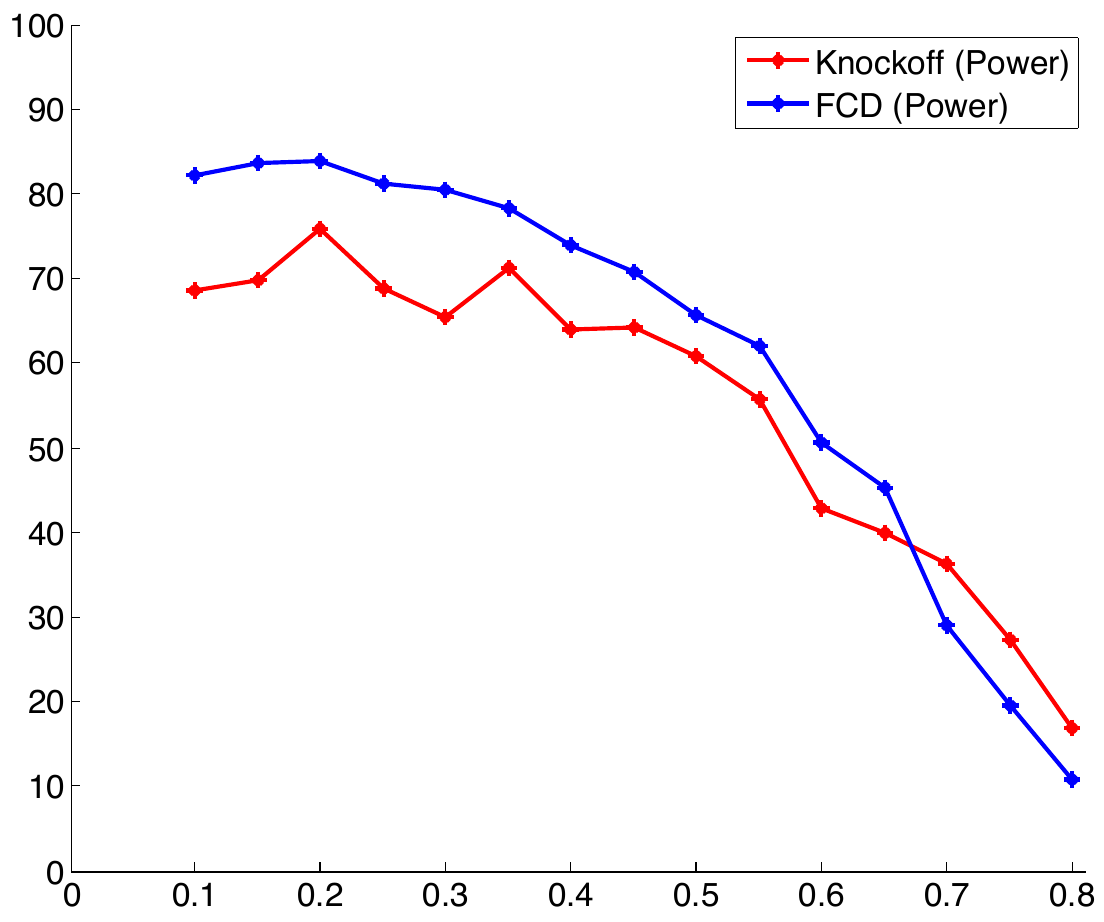}
\end{tabular}
\put(-305,-62){{\scriptsize Feature Correlation $(\eta)$}}
\put(-110,-62){{\scriptsize Feature Correlation $(\eta)$}}
\put(-345,-10){\rotatebox{90}{{\scriptsize $\FDRdir$}}}
\put(-150,-25){\rotatebox{90}{{\scriptsize statistical power}}}

\caption{Testing FCD and model free knockoff methods with varying the feature correlation parameter $\eta$. Here, $n = 700$, $p = 1000$, $k = 50$, $A = 4.5$. The target level is $q = 10\%$. $\FDRdir$ and power are computed by averaging over $100$ realizations of noise and design matrices.}\label{fig:rho}
\end{figure}

\noindent{{\bf Effect of feature correlation:}}
We test the effect of feature correlations on the performance of FCD procedure, comparing it with the model free knockoff. We set $n =700$, $p = 1000$, $k = 50$, $A = 4.5$. Recall that the rows of the design matrix $X$ are generated from a $\normal(0,\Sigma)$ distribution, with $\Sigma_{ij} = \eta^{|i-j|}$, and then the columns of $X$ are normalized to have unit norm. We vary the  parameter $\eta$ in the set $\{0.1, 0.15, 0.2, \dotsc, 0.75, 0.8\}$. For each value of $\eta$, we compute $\FDRdir$ and power for both methods, by averaging over $100$ realizations of noise and design matrix $X$. The results are displayed in Figure~\ref{fig:rho}. 

As observed, both methods control $\FDRdir$ over the range of correlations tested. From the power plot, we see that the power of both methods decays as the features  correlations increase. This is expected because when the features are highly correlated it is harder to distinguish between them and report the truly significant ones. Indeed, for large values of $\eta$, both methods select a few variables. This way, $\FDRdir$ is still controlled but the power is low.
The proposed FCD procedure has higher power than model free knockoff for $\eta\le 0.65$.

\bigskip

\noindent{{\bf Effect of Sparsity:}}
Here, we set $n = 2000$, $p = 3000$, $A = 4.5$, $\eta = 0.1$ and vary the sparsity level of the coefficients in the set $k\in\{10,15, 20, \dotsc, 130\}$. For both methods, the power and FDR are computed by averaging over $100$ trials of noise and the generation of coefficient vector $\theta_0$.
Both methods control $\FDRdir$ over the entire range, with FCD achieving lower $\FDRdir$ for small values of $k$. In terms of power, both methods have close power, and the FCD has higher power for small $k$. 

\begin{figure}[]

\begin{tabular}{lll}
\includegraphics[width = 4.8cm]{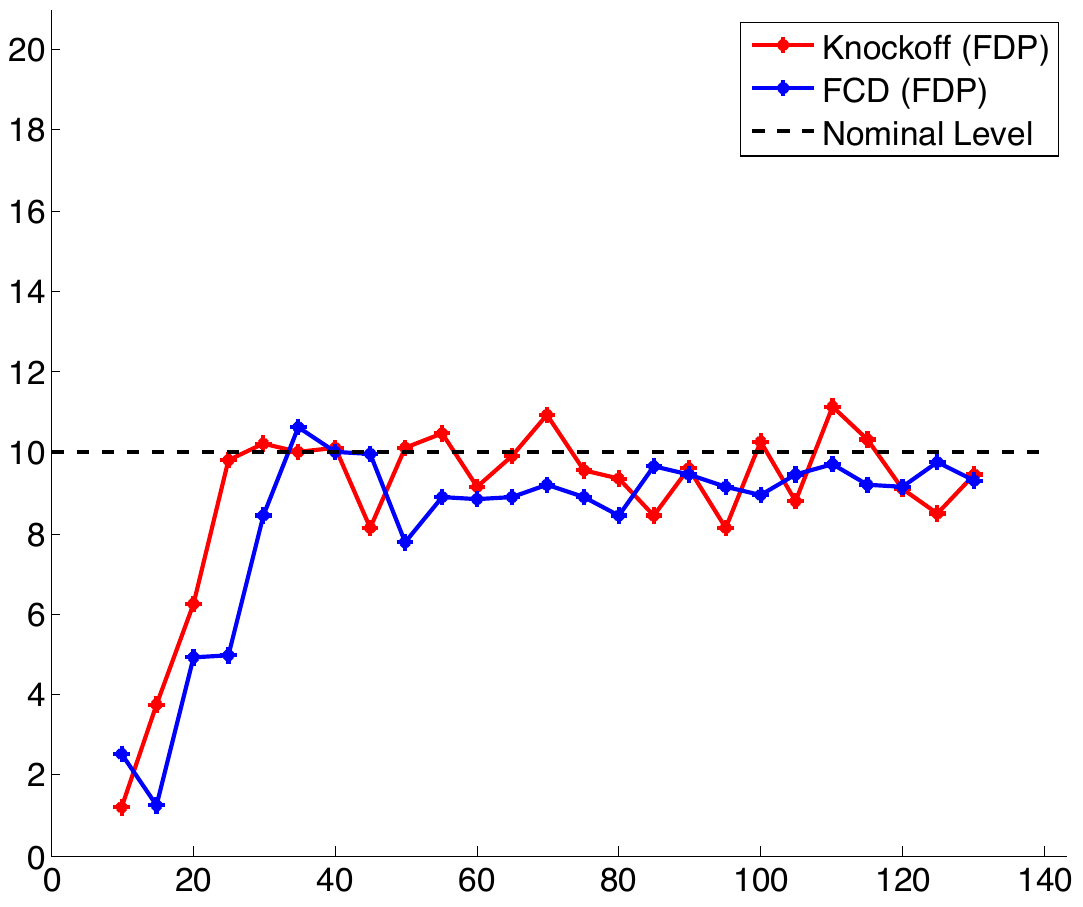}&\phantom{AAAAA}&
\includegraphics[width=4.8cm]{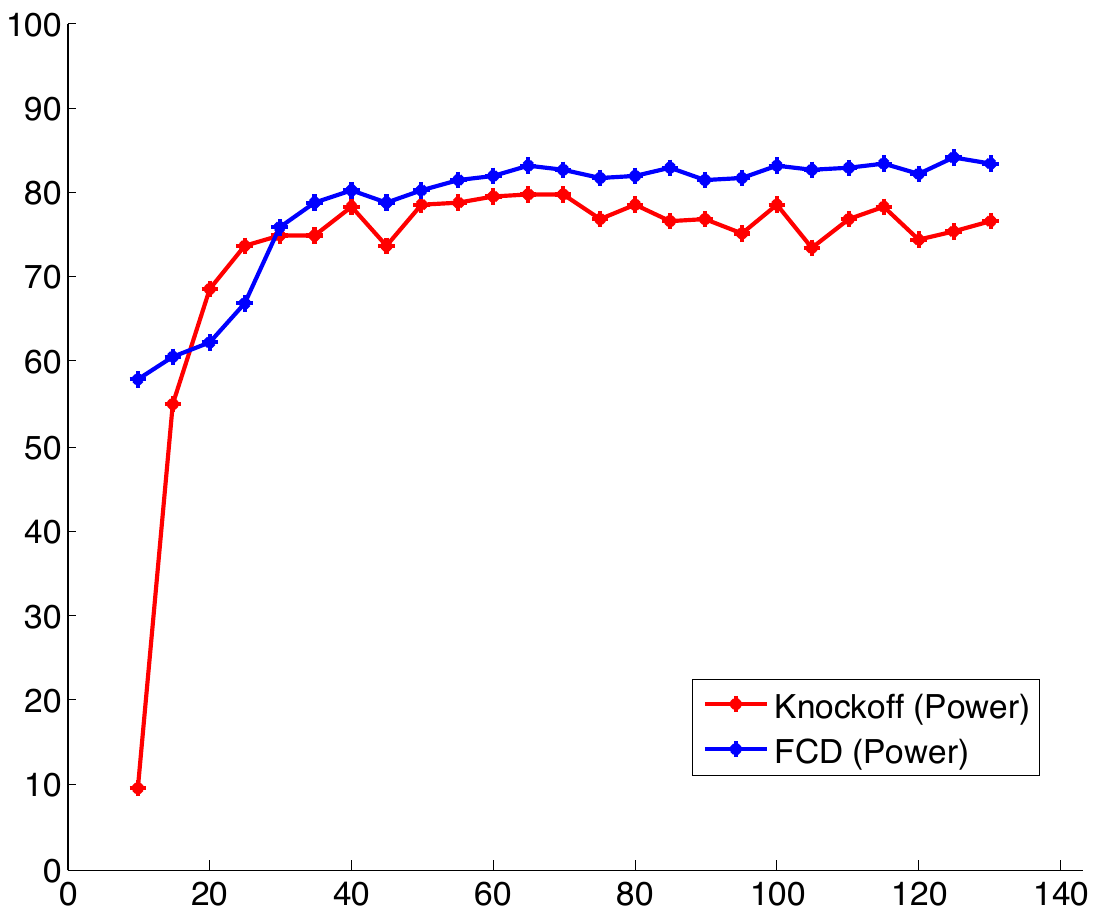}
\end{tabular}
\put(-297,-62){{\scriptsize Sparsity Level $(k)$}}
\put(-103,-62){{\scriptsize Sparsity Level $(k)$}}
\put(-345,-10){\rotatebox{90}{{\scriptsize $\FDRdir$}}}
\put(-150,-25){\rotatebox{90}{{\scriptsize statistical power}}}

\caption{Testing FCD and model free knockoff methods with varying the sparsity level $k$. Here, $n = 2000$, $p = 3000$, $A = 4.5$, $\eta = 0.1$. The target level is $q = 10\%$. $\FDRdir$ and power are computed by averaging over $100$ realizations of noise and coefficient vectors.}\label{fig:rho_sparsity_level}
\end{figure}

\section{Real Data Experiments}\label{sec:realdata}
In this section we evaluate the proposed method to find the mutations in the Human Immunodeficiency Virus Type 1 
associated with drug resistance\footnote{The dataset is available online at \url{https://hivdb.stanford.edu/pages/published_analysis/genophenoPNAS2006}.}. This dataset is presented and analyzed in \cite{rhee2006genotypic} and is obtained by 
analyzing HIV-1 subtype B sequences from persons with histories of antiretroviral treatment.
The dataset contains the mutations in the protease
and reverse transcriptase (RT) positions of the HIV-1 subtype B sequences which correspond
to resistance to Protease Inhibitors (PI), to nucleoside reverse transcriptase inhibitors (NRTIs) and to non-nucleoside RT inhibitors (NNRTIs).

 In order to deal with missing measurements and preprocessing the dataset we mostly follow the steps taken in \cite{barber2015controlling}. The design matrix $X \in \{0,1\}^{n\times p}$ is formed by letting $X_{ij} = 1$ if the $i$'th sample
contains the $j$'th mutation and $X_{ij} = 0$ otherwise. Further, for a specific drug, 
the $i$'th entry of the response vector $y_i$ denotes the 
logarithm of the increase in the resistance to that drug in the $i$'th patient. We let $q = 0.2$ and we apply the FCD procedure
described in subsection \ref{subsec:FCD} to detect the mutations in the HIV-1 associated with resistance to each drug. In order 
to evaluate the performance of our method, we compare it with the knockoff filter procedure \cite{barber2015controlling}
with the test statistics based on lasso. The size of the dataset ($n$, $p$) for each drug is noted under the bar plot corresponding to that drug. For all cases, except the data for resistance to TDF, we have $n>2p$. 

We have 
used two different methods for generating the knockoff variables; in knockoff1, the knockoff variables are generated by solving 
a semi-definite program (SDP) and in knockoff2, equi-correlated knockoff variables are created without solving an SDP at 
a lower computational cost\footnote{More information regarding the procedure are available at \url{https://web.stanford.edu/group candes/knockoffs/software/knockoff/index.html}.}. Since this is a real data experiment, there is no ground truth. However, 
we use the methodology in \cite{barber2015controlling} to assess our results. In order to do this, we evaluate the reproducibility of 
the outcomes of these procedures by comparing them with treatment-selected mutation (TSM) panels provided 
in \cite{rhee2005hiv}. These panels contain mutations that are observed more frequently in virus samples from patients 
that have been treated by each drug in compare with the patients who have never been treated with that drug. Since 
these panels are created independently from the dataset that we use, they can provide a good measure 
for validating the reproducibility of the results obtained by each procedure.  

\begin{figure}
\begin{center}
\includegraphics[width=4.5in]{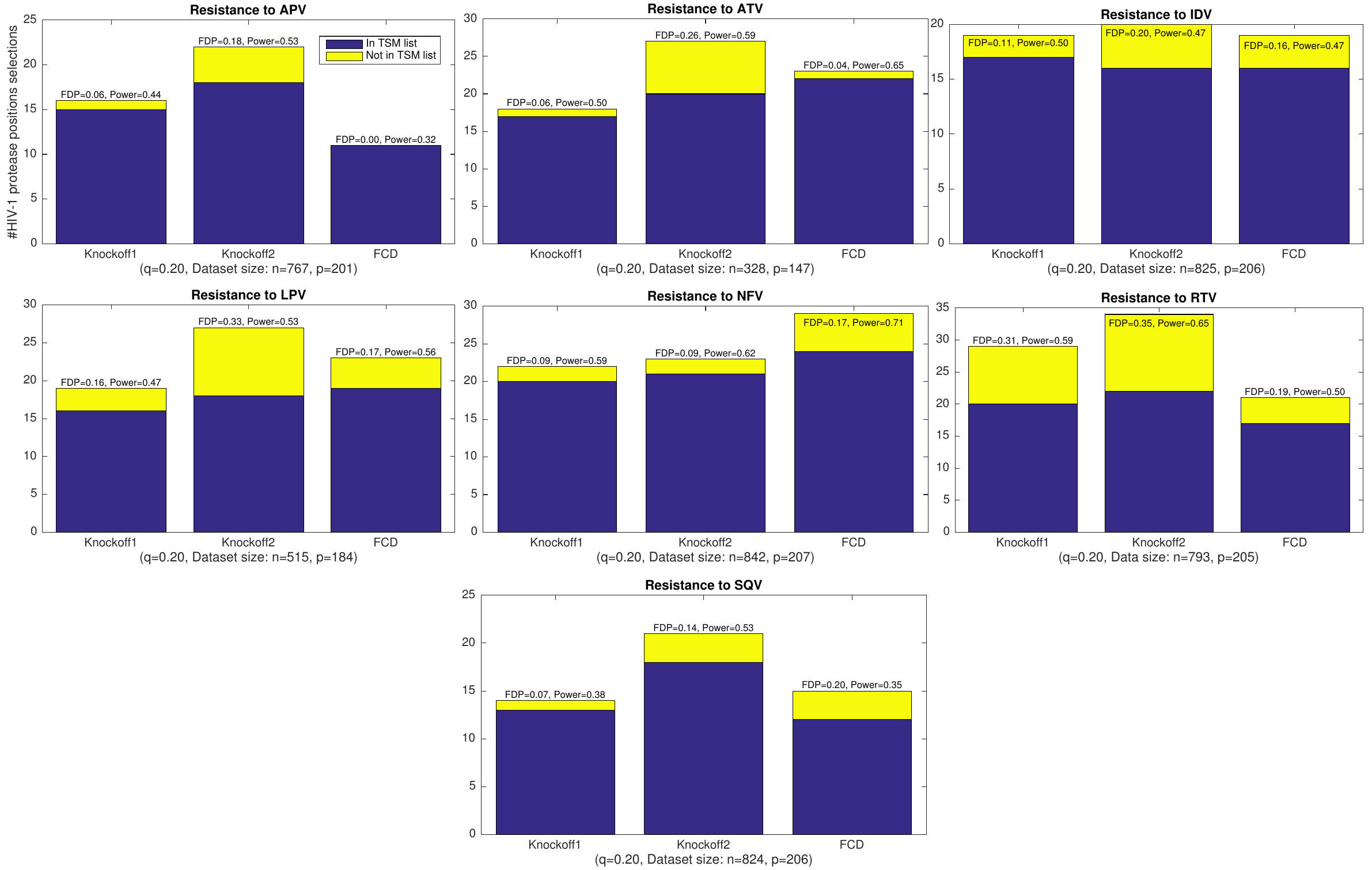}

\end{center}
    \caption{Summary of the results of applying the knockoff filter and FCD for detecting the
    mutation positions in HIV-1 associated with resistance to type-PI drugs using the dataset in \cite{rhee2006genotypic}. 
    In these experiments we have used $q=0.2$.
    In the plots, blue bars show the number of detected positions by different methods that appear in the TSM panels. On top of
    each bar the proportion of detected mutations that appear in the TSM panel (an estimate for FDP) and the 
    proportion of mutations in the TSM panel that are detected by different methods (an estimate for power) are stated.}
    \label{fig:PI_drugs}
\end{figure}

\begin{figure}
\begin{center}
\includegraphics[width = 4.5in]{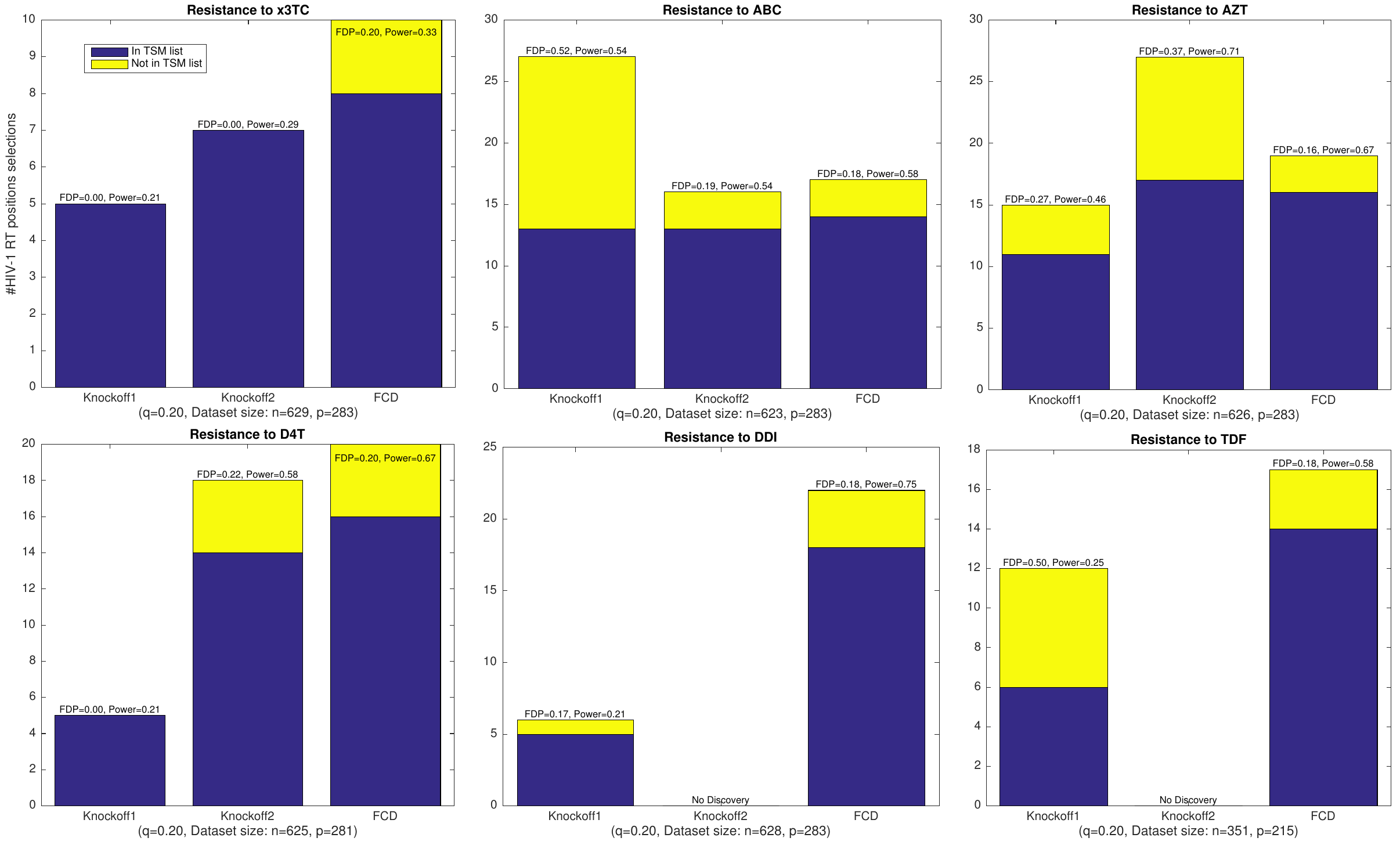}
\end{center}
    \caption{Same as Figure \ref{fig:PI_drugs} for type-NRTI drugs.}
    \label{fig:NRTI_drugs}
\end{figure}

\begin{figure}
\begin{center}
\includegraphics[width=4.5in]{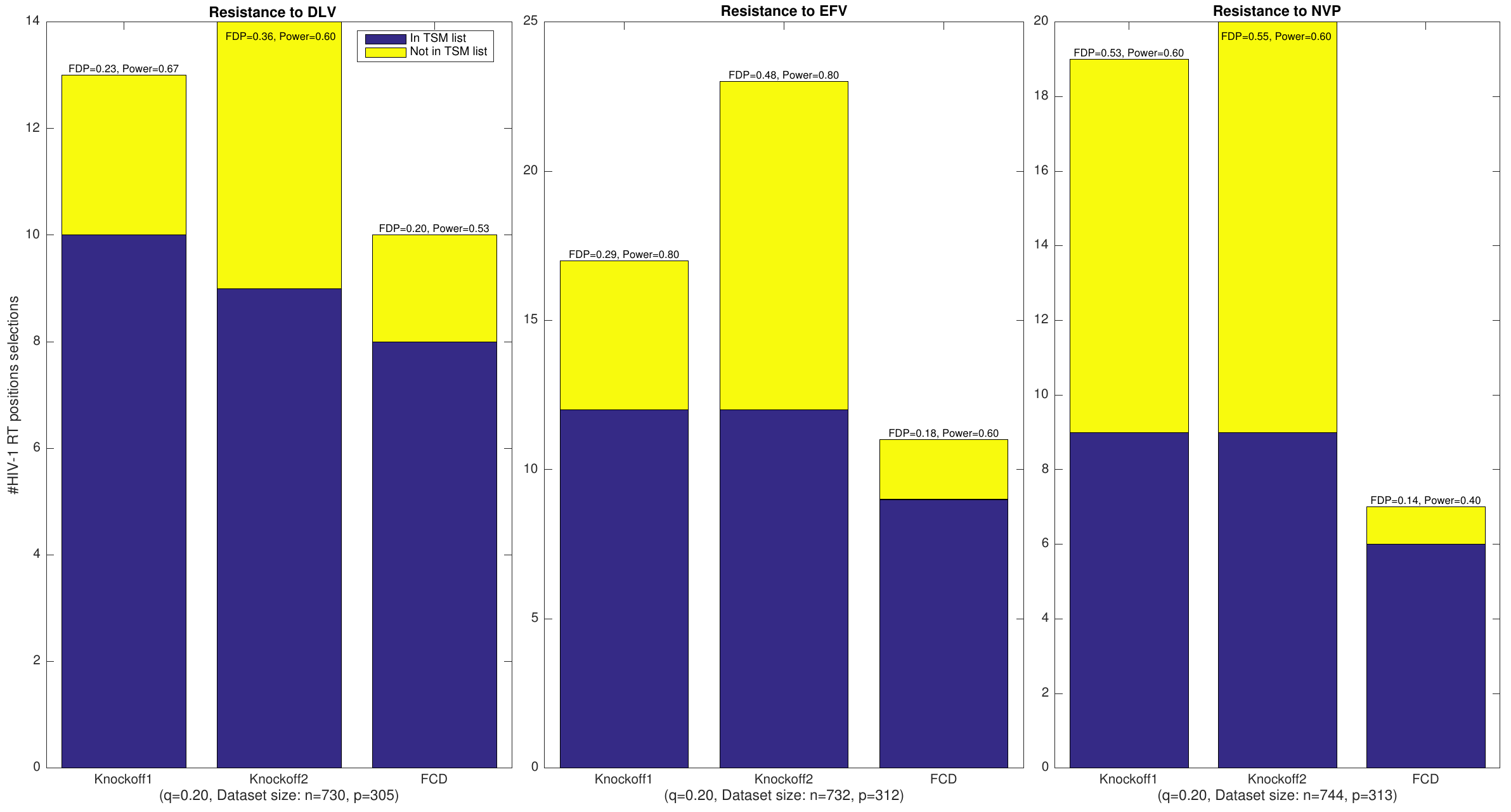}
\end{center}

    \caption{Same as Figure \ref{fig:PI_drugs} for type-NNRTI drugs.}
    \label{fig:NNRTI_drugs}
\end{figure}

A summary of the results can be seen in Figures \ref{fig:PI_drugs}, \ref{fig:NRTI_drugs}, \ref{fig:NNRTI_drugs}. 
It can be seen that the FCD method achieves the target FDR level of $q=0.2$ and the obtained power in half of 
the cases (8 out of 16 drugs) is larger than the power achieved by the knockoff filter. Overall, the achieved power is 
comparable with the power of the knockoff filter method. 

\section{Proof of Main Theorems}
\subsection{Proof of Theorem~\ref{thm:main}}
Define $G(t) = 2(1-\Phi(t))$, with $\Phi(t)$ denoting the standard Gaussian cumulative distribution. We start by two lemmas about the properties of $G(t)$.
\begin{lemma}\label{lem:G1}
For all $t\ge 0$, we have 
\begin{align}
\frac{2}{t+1/t} \phi(t)< G(t) <\frac{2}{t}\phi(t)\,,
\end{align}
where $\phi(t) = e^{-t^2/2}/\sqrt{2\pi}$ is the standard Gaussian density.
\end{lemma}
Lemma~\ref{lem:G1} is the standard trial bound on the Gaussian distribution and its proof is omitted.
\begin{lemma}\label{lem:G2}
For all $t>0$, $\eps<\min(1,1/t)$ and $\delta<\min(1,1/t^2)$, the following holds true:
\begin{align}
\frac{G((1-\delta)t - \eps)}{G(t)}\le 1+ 8 (\eps+\eps t +\delta + \delta t^2)\,.
\end{align}
\end{lemma}
Proof of Lemma~\ref{lem:G2} is given in Appendix~\ref{proof:lem-G2}.

Using Proposition~\eqref{propo:biasvardecompose}, we have
\begin{align}
\label{eq:Tdecompose-b}
T_i = \frac{\sqrt{n}\theta_{0,i}}{\hsigma\sqrt{\Lambda_{ii}}} + \frac{\sigma}{\hsigma} \tZ_i + \frac{\Delta_i}{\hsigma\sqrt{\Lambda_{ii}}}
\end{align}
where $\tZ_i\sim\normal(0,\Lambda^0)$. 
By invoking \cite[Lemma 3.1]{javanmard2014confidence}, $\Lambda_{ii}$ are bounded from below by an arbitrary fixed constant $0<c<1$, for large enough $n$.
In addition, since $\left|\Lambda^0 - \Omega^0\right|_\infty = o_p(1)$, for $(i,j) \in \Gamma(\gamma,c_0)^c$ we have 
\begin{align}
|\Lambda^0_{ij}| < C(\log p)^{-2-\gamma}\,,
\end{align}
for some constant $C >0$. Further, by Condition $(iii)$ in the theorem statement, we have 
\begin{align}
\label{eq:resasump2-b}
\Big|\Big\{(i,j): |\Omega_{ij}^0| > \frac{1-\rho}{1+\rho}\Big\}\Big| = O(p)\,.
\end{align}
Define $\pset \equiv \{i\in [p]: \theta_{0,i} \ge 0\}$ and $\nset \equiv \{i\in [p]: \theta_{0,i} \le 0\}$. 

We first consider the case that $t_0$, given by~\eqref{t0-1}, does not exist.
In this case, $t_0  = \sqrt{2\log p}$ and for any $\eps>0$ we have 
\begin{align}
\prob\Bigg(\sum_{i} \ind\left(\hsign_i \neq \sign(\theta_{0,i})\right) \ge 1 \Bigg) &\le
\prob\Bigg(\sum_{i\in\nset} \ind\left(T_i\ge \sqrt{2\log p}\right) \ge 1 \Bigg) \\
&+ \prob\Bigg(\sum_{i\in\pset} \ind\left(T_i\leq -\sqrt{2\log p}\right) \ge 1 \Bigg).
\end{align}

We can bound the first term on the right hand side above as 
\begin{align*}
\prob\Bigg(\sum_{i\in \nset} \ind\left(T_i\ge \sqrt{2\log p}\right) \ge 1 \Bigg) \\
&\hspace{-0.5in}\leq 
\prob\Bigg(\sum_{i\in \nset} \ind\left(\frac{\sigma}{\hsigma}\tZ_i + \frac{\Delta_i}{\hsigma\sqrt{\Lambda_{ii}}}\ge \sqrt{2\log p}\right) \ge 1 \Bigg)\\
&\hspace{-0.5in}\leq\prob\Bigg(\sum_{i\in \nset} \ind\left(\tZ_i \geq \frac{\hsigma}{\sigma}\sqrt{2\log p} - \frac{\left\|\Delta\right\|_\infty}{\sigma\sqrt{c}}\right) \ge 1 \Bigg)\\
&\hspace{-0.5in}\leq p \max_{i \in [p]} \prob \left(\tZ_i \geq (1-\eps)\sqrt{2\log p} - \eps\right) \\
&\hspace{-0.5in}+ \prob\left\{\left\|\Delta\right\|_\infty \geq \sigma\eps\sqrt{c}\right\} + \prob\left\{\left|\frac{\hsigma}{\sigma}-1\right| \geq \eps\right\}\\
&\hspace{-0.5in}\leq \frac{p}{2}G\left((1-\eps)\sqrt{2\log p} - \eps\right) \\
&\hspace{-0.5in}+  \prob\left\{\left\|\Delta\right\|_\infty \geq \sigma\eps\sqrt{c}\right\} + \prob\left\{\left|\frac{\hsigma}{\sigma}-1\right| \geq \eps\right\}\,.
\end{align*}

which goes to zero as $n, p \to \infty$, due to Proposition~\ref{propo:biasvardecompose} along with Condition $(i)$, that $s_0 = o(\sqrt{n}/(\log p)^2)$, and using Lemma~\ref{lemma:sigmahatoversigma}. Similarly, and by symmetry, the second term goes to zero as $n, p  \to \infty$ and the claim follows. 

We next focus on the event that $t_0$, given by~\eqref{t0-1} exists. By definition of $t_0$ in this case, we have 
\begin{align*}
\frac{p G(t_0)}{R(t_0)\vee 1} = q\,.
\end{align*}
(Indeed, it is clear that the left-hand side is at most $q$. Equality holds since $t_0$ is the minimum $t$, with such property.)

Define $Q(t)\equiv G(t)/2$ for all $t\in \reals$. Let 

\begin{align}
 A_p &\equiv \sup_{0\leq t\leq t_p}\left|\frac{\sum_{i\in\pset}\left\{\ind\left(T_i \leq - t\right) - Q(t) \right\} + \sum_{i\in\nset}\left\{\ind\left(T_i \geq t\right) - Q(t) \right\} }{pG(t)}\right|
\end{align}

Then,
\begin{align}
\dFDP(t_0) &= \frac{\sum_{i\in\pset}\ind(T_i\le-t_0) + \sum_{i\in\nset} \ind(T_i\ge t_0) }{R(t_0)\vee1} \nonumber\\
&\leq \frac{pG(t_0)A_p + s_0 Q(t_0) + 2(p-s_0)Q(t_0)}{R(t_0)} \\
&\leq\frac{pG(t_0) (1+A_p)}{R(t_0)} \le q(1+ A_p).
\end{align}
Hence, we need to prove that $ A_p\to 0$, in probability. Note that 

\begin{align}
\label{eq:apineq}
A_p &\leq \sup_{0\leq t\leq t_p}\Bigg\{\left|\frac{\sum_{i\in \nset}\left\{\ind\left(T_i \geq t\right) - Q(t) \right\}}{pG(t)}\right|\\
&\quad\quad\quad\quad\quad+ \left|\frac{\sum_{i\in \pset}\left\{\ind\left(T_i \leq - t\right) - Q(t) \right\}}{pG(t)}\right| \Bigg\}\nonumber\\
&\leq \sup_{0\leq t\leq t_p}\left|\frac{\sum_{i\in\nset}\left\{\ind\left(T_i \geq t\right) - Q(t) \right\}}{pG(t)}\right| \\
&+  \sup_{0\leq t\leq t_p}\left|\frac{\sum_{i\in\pset}\left\{\ind\left(T_i \leq - t\right) - Q(t) \right\}}{pG(t)}\right| \,.
\end{align}

Note that by symmetry it is sufficient to prove that the first term in \eqref{eq:apineq} goes to zero in probability.
Consider a discretization $0\le \tau_1 < \tau_2< \dotsc <\tau_b = t_p$ such that $\tau_j - \tau_{j-1} = v_p$, for
$1\le j\le b-1$ and $\tau_b - \tau_{b-1} \le v_p$, where $v_p = 1/\sqrt{\log p}$. Hence, $b\sim t_p/v_p$. For any $t\in [\tau_{j-1}, \tau_j]$, we have
\begin{align*}
\frac{\sum_{i\in\nset} \ind(T_i \ge \tau_j)}{p Q(\tau_j)} \cdot \frac{Q(\tau_j)}{Q(\tau_{j-1})} &\le
\frac{\sum_{i\in\nset} \ind(T_i \ge t)}{p Q(t)} \\
&\le
\frac{\sum_{i\in\nset} \ind(T_i \ge \tau_{j-1})}{p Q(\tau_{j-1})}\cdot \frac{Q(\tau_{j-1})}{Q(\tau_{j})}
\end{align*}
Hence, it suffices to show that
\begin{align}
\max_{0\le j\le b}\, \bigg|\frac{\sum_{i\in\nset} \{\ind(T_i \ge \tau_j) - Q(\tau_j)\} }{p Q(\tau_j)} \bigg| \to 0\,\label{eq:inter1b}
\end{align}
in probability. 

In the following lemma, we provide sufficient conditions to obtain Eq.~\eqref{eq:inter1b}.
\begin{lemma}\label{lem:inter1}
Suppose that for any $\delta>0$, the followings hold:
\begin{align}
&\sup_{0\le t\le t_p} \prob \bigg\{\bigg|\frac{\sum_{i\in\nset}  \ind(T_i\ge t) }{p Q(t)} - 1 \bigg|\ge \delta \bigg\} = o(1)\label{sup-vpb}
\end{align}
and
\begin{align}
&\int_{0}^{t_p}\prob \bigg\{\bigg|\frac{\sum_{i\in\nset}  \ind(T_i\ge t) }{p Q(t)} - 1 \bigg|\ge \delta \bigg\} \de t= o(v_p)
\label{int-vpb}
\end{align}
where $t_p = (2\log p - 2\log\log p)^{1/2}$ and $v_p = (\log p)^{-1/2}$, then \eqref{eq:inter1b} hold true.
\end{lemma}
We refer to Appendix~\ref{proof:lem-inter1} for the proof of Lemma~\ref{lem:inter1}.

By virtue of Lemma~\ref{lem:inter1} we only need to prove Eqs.~\eqref{sup-vpb} and \eqref{int-vpb}. We start by analyzing the following expression

\begin{align}
\label{eq:expected-sumoverpairs-a}
\E \bigg\{\bigg|\frac{\sum_{i\in\nset}\{\ind(T_i\ge t) - \prob(T_i\ge t)\} }{p Q(t)} \bigg|^2 \bigg\} \\
&\hspace{-1in}\le 
\frac{\sum_{i,j\in\nset} \{\prob(T_i\ge t, T_j \ge t) - \prob(T_i \ge t) \prob(T_j \ge t)\}}{p_0^2 Q(t)^2}\nonumber\\
&\hspace{-1in}\le \frac{1}{p_0^2}\sum_{i,j\in\nset}\frac{\prob(T_i\ge t, T_j \ge t)}{Q(t)^2} - 1\nonumber\\
&\hspace{-1in}\le \frac{1}{p_0^2}\sum_{i,j\in[p]}\frac{\prob(\tT_i\ge t, \tT_j \ge t)}{Q(t)^2} - 1,
\end{align} 

with $p_0 = |\nset|$ and 
\begin{align}
\tT_i \equiv \frac{\sigma}{\hsigma} \tZ_i + \frac{\Delta_i}{\hsigma\sqrt{\Lambda_{ii}}}\,.
\end{align}
The last inequality of~\eqref{eq:expected-sumoverpairs-a} holds because $\theta_{0,i} \le 0$ for $i\in \nset$ and therefore $T_i\le \tT_i$ (Recall definition of $T_i$, given by Eq.~\eqref{eq:Tdecompose-b}.) Further, because $S^c = \{i\in [p]: \theta_{0,i} = 0\}\subseteq \nset$, we have $p_0\ge p-s_0$. Since $s_0 = o(\sqrt{n}/(\log p)^2)$ by Condition ($i$), we have $p_0 = \Omega(p)$. 

We partition the set $\{(i,j): i, j\in [p]\}$ into two disjoint sets, namely $\Gamma(\gamma,c_0)$ (highly correlated test statistics) and $\Gamma(\gamma,c_0)^c$ (weakly correlated test statistics). (Recall the definition of set $\Gamma(\gamma,c_0)$ given by~\eqref{eq:Lambda(gamma)}.) We analyze the contribution of each set separately. 

\subsubsection{Highly correlated test statistics ($\Gamma(\gamma,c_0)$)}

We first consider the set $\Gamma(\gamma,c_0)$. Note that $(\tZ_i, \tZ_j)\sim \normal\left(0, \begin{bmatrix}
1 & \Lambda_{ij}^0\\
\Lambda_{ij}^0 & 1\end{bmatrix}\right)$. Using 
\eqref{eq:Tdecompose-b}, we have
\begin{align}
\prob\left(\tT_i\ge t, \tT_j \ge t\right)  &\le \prob\left(\tZ_i >\frac{\hsigma}{\sigma}t - \frac{\Delta_i}{\sigma\sqrt{\Lambda_{ii}}}, \tZ_j>\frac{\hsigma}{\sigma}t - \frac{\Delta_j}{\sigma\sqrt{\Lambda_{jj}}}\right) \nonumber\\
&\le \prob\left(\tZ_i>(1-\eps_1)t -\eps_2, \tZ_j>(1-\eps_1)t - \eps_2\right)\nonumber \\
&+  \prob\left\{\left\|\Delta\right\|_\infty \geq \sigma\eps_2\sqrt{c}\right\} + \prob\left\{\left|\frac{\hsigma}{\sigma}-1\right| \geq \eps_1\right\}\label{eq:TiTjboundprem}\\
&\le C\left((1-\eps_1)t - \eps_2+1\right)^{-2} \exp\left\{-\frac{\left((1-\eps_1)t - \eps_2\right)^2}{1+\Lambda^0_{ij}}\right\} \nonumber\\
&+  \prob\left\{\left\|\Delta\right\|_\infty \geq \sigma\eps_2\sqrt{c}\right\} + \prob\left\{\left|\frac{\hsigma}{\sigma}-1\right| \geq \eps_1\right\},\label{eq:TiTjbound}
\end{align} 
where the last inequality follows from~\cite[Lemma 6.2]{liu2013gaussian}.

Let $\Psi(\rho) \equiv  \Big\{(i,j):  i, j\in[p], |\Lambda^0_{ij}| > (1-\rho)/(1+\rho) \Big\}$. Note that 
by \eqref{eq:resasump2-b} and since $\left|\Lambda^0 - \Omega^0\right|_\infty = o_p(1)$, we have $|\Psi(\rho)| = O(p)$. We can write
\begin{align}
\label{eq:Psifrag}
\frac{1}{p_0^2} \sum_{(i,j)\in \Gamma(\gamma,c_0)}\frac{\prob(\tT_i\ge t, \tT_j \ge t)}{Q(t)^2} \le
\frac{1}{p_0^2}&\bigg[\sum_{(i,j)\in \Psi(\rho)}\frac{\prob(\tT_i\ge t, \tT_j \ge t)}{Q(t)^2} \nonumber\\
&\;\;+ \sum_{(i,j) \in \Gamma(\gamma,c_0)\setminus \Psi(\rho)}\frac{\prob(\tT_i\ge t, \tT_j \ge t)}{Q(t)^2}\bigg].
\end{align}
We treat the terms on the right hand side separately. For the first term, since $\Lambda^0_{ij} \le 1$, by
using \eqref{eq:TiTjbound}, we have

\begin{align*}
\frac{1}{p_0^2}\sum_{(i,j)\in \Psi(\rho)}\frac{\prob(\tT_i\ge t, \tT_j \ge t)}{Q(t)^2} \\
&\hspace{-1in}\leq \frac{|\Psi(\rho)|}{p_0^2Q(t)^2}\Bigg\{C\left((1-\eps_1)t - \eps_2\right)^{-2} \exp\left(-\left((1-\eps_1)t - \eps_2\right)^2/2\right)\nonumber\\
&\hspace{-0.28in}+\prob\left\{\left\|\Delta\right\|_\infty \geq \sigma\eps_2\sqrt{c}\right\} + \prob\left\{\left|\frac{\hsigma}{\sigma}-1\right| \geq \eps_1\right\}\Bigg\}\nonumber\\
&\hspace{-1in}\le Cp^{-1} \Bigg\{\left(\frac{G((1-\eps_1)t-\eps_2)}{G(t)}\right)^2\exp(((1-\eps_1)t-\eps_2)^2/2) \nonumber\\
&\hspace{-0.48in}+\frac{1}{Q(t)^2}\left(\prob\left\{\left\|\Delta\right\|_\infty \geq \sigma\eps_2\sqrt{c}\right\} + \prob\left\{\left|\frac{\hsigma}{\sigma}-1\right| \geq \eps_1\right\}\right)\Bigg\},
\end{align*}

where in the second inequality we used the fact that $|\Psi(\rho)| = O(p)$ and that $p_0 = \Omega(p)$.
Take $\eps_2 = s_0(\log p)/\sqrt{n}$. 
By using Lemmas \ref{propo:biasvardecompose}, \ref{lemma:sigmahatoversigma}, and~\ref{lem:G2}, we get

\begin{align}
\frac{1}{p_0^2} \sum_{(i,j)\in \Psi(\rho)}\frac{\prob(\tT_i\ge t, \tT_j \ge t)}{Q(t)^2} &\le C p^{-1}\left(1+\eps_1^2+ \eps_2^2+\eps_2^2 t^2 + \eps_1^2 t^4 \right)e^{t^2/2} \nonumber\\
&+ C\frac{p^{-1}}{Q(t)^2}\bigg(\prob\left\{\left\|\Delta\right\|_\infty \geq \sigma\eps_2\sqrt{c}\right\} \\
&\quad\quad\quad\quad\quad+ \prob\left\{\left|\frac{\hsigma}{\sigma}-1\right| \geq \eps_1\right\}\bigg)\nonumber\\
&\leq  C p^{-1}\bigg\{(1+\eps_1^2+\eps_2^2)e^{t^2/2}\\
&\hspace{0.6in}+e^{t^2/2}\eps_2^2t^2 + e^{t^2/2}\eps_1^2t^4 \nonumber\\
&+ \frac{1}{Q(t)^2}\left(e^{-c_1 n} + p^{-c_2} +\prob\left\{\left|\frac{\hsigma}{\sigma}-1\right| \geq \eps_1\right\} \right)\bigg\}\,\label{gamma-B11}\,,
\end{align}

for some constant $C>0$. 

To bound the second term on the right-hand side of Eq.~\eqref{eq:Psifrag}, note that for $(i,j) \in \Gamma(\gamma,c_0)\setminus\Psi(\rho)$, we have $\Lambda_{ij}^0 \leq (1-\rho)/(1+\rho)$. Thus,
using \eqref{eq:TiTjbound}

\begin{align*}
\frac{1}{p_0^2} \sum_{(i,j)\in \Gamma(\gamma,c_0)\setminus\Psi(\rho)}\frac{\prob(\tT_i\ge t, \tT_j \ge t)}{Q(t)^2} \\
&\hspace{-1.6in}\le \frac{|\Gamma(\gamma,c_0)|}{p_0^2Q(t)^2}\Bigg\{C\left((1-\eps_1)t - \eps_2\right)^{-2} \exp\left(-(1+\rho)\left((1-\eps_1)t - \eps_2\right)^2/2\right)\nonumber\\
&\hspace{-0.85in}+\prob\left\{\left\|\Delta\right\|_\infty \geq \sigma\eps_2\sqrt{c}\right\} + \prob\left\{\left|\frac{\hsigma}{\sigma}-1\right| \geq \eps_1\right\}\Bigg\}\nonumber\\
&\hspace{-1.6in}\le c'p^{\rho - 1} \Bigg\{\left(\frac{G((1-\eps_1)t-\eps_2)}{G(t)}\right)^2\exp\left((1-\rho)\left((1-\eps_1)t-\eps_2\right)^2/2\right) \nonumber\\
&\hspace{-1.22in}+\frac{1}{Q(t)^2}\left(\prob\left\{\left\|\Delta\right\|_\infty \geq \sigma\eps_2\sqrt{c}\right\} + \prob\left\{\left|\frac{\hsigma}{\sigma}-1\right| \geq \eps_1\right\}\right)\Bigg\},
\end{align*}

for any arbitrary constant $c'>0$.

Hence, using Lemmas \ref{propo:biasvardecompose}, \ref{lemma:sigmahatoversigma}, and~\ref{lem:G2}, we get

\begin{align}
\frac{1}{p_0^2} \sum_{(i,j)\in \Gamma(\gamma,c_0)\setminus \Psi(\rho)}\frac{\prob(\tT_i\ge t, \tT_j \ge t)}{Q(t)^2} \\
&\hspace{-1.5in}\le c' p^{\rho - 1}\left(1+\eps_1^2+\eps_2^2+\eps_2^2 t^2 + \eps_1^2 t^4 \right)e^{(1-\rho)t^2/2} \nonumber\\
&\hspace{-1.5in}+ c'\frac{p^{\rho-1}}{Q(t)^2}\left(\prob\left\{\left\|\Delta\right\|_\infty \geq \sigma\eps_2\sqrt{c}\right\} + \prob\left\{\left|\frac{\hsigma}{\sigma}-1\right| \geq \eps_1\right\}\right)\nonumber\\
&\hspace{-1.5in}\leq  c' p^{\rho-1}\bigg\{(1+\eps_1^2+\eps_2^2) e^{(1-\rho)t^2/2}+e^{(1-\rho)t^2/2}\eps_2^2t^2 + e^{(1-\rho)t^2/2}\eps_1^2t^4 \nonumber\\
&\hspace{-0.9in} + \frac{1}{Q(t)^2}\left(e^{-c_1 n} + p^{-c_2} +\prob\left\{\left|\frac{\hsigma}{\sigma}-1\right| \geq \eps_1\right\} \right)\bigg\}\,\label{gamma-B12}
\end{align}

uniformly for $0\le t\le t_p$.
and for any arbitrary constant $c'>0$.
 
\subsubsection{Weakly correlated test statistics ($\Gamma(\gamma,c_0)^c$)} 
We next consider $\Gamma(\gamma,c_0)^c\cap\nset$. Using~\cite[Lemma 6.1]{liu2013gaussian} (for $d=2$ in its statement) and Eq.~\eqref{eq:TiTjboundprem}, we have
\begin{align}
&\sup_{0\le t\le t_p} \bigg| \frac{\prob(\tT_i\ge t, \tT_j\ge t)}{Q(t)^2} - 1\bigg|\nonumber\\
&\le \sup_{0\le t\le t_p} \bigg|\frac{1}{Q(t)^2} \prob\left(\tZ_i >\frac{\hsigma}{\sigma}t - \frac{\Delta_i}{\sigma\sqrt{\Lambda_{ii}}}, \tZ_j>\frac{\hsigma}{\sigma}t - \frac{\Delta_j}{\sigma\sqrt{\Lambda_{jj}}}\right)-1\bigg| \nonumber\\
&\le \sup_{0\le t\le t_p} \bigg| \frac{1}{Q(t)^2}{\prob\left(\tZ_i>(1-\eps_1)t -\eps_2, \tZ_j>(1-\eps_1)t -\eps_2)\right)} - 1\bigg|\nonumber\\
&+ \frac{1}{Q(t_p)^2}\left(\prob\left\{\left\|\Delta\right\|_\infty \geq \sigma\eps_2\sqrt{c}\right\} + \prob\left\{\left|\frac{\hsigma}{\sigma}-1\right| \geq \eps_1\right\}\right)\nonumber\\
&\le\sup_{0\le t\le t_p}\left(\frac{Q((1-\eps_1)t-\eps_2)}{Q(t)}\right)^2\\
&\quad\sup_{0\le t\le t_p} \left| \frac{\prob\left(\tZ_i>(1-\eps_1)t -\eps_2, \tZ_j>(1-\eps_1)t -\eps_2)\right)}{Q((1-\eps_1)t-\eps_2)^2} - 1\right| \nonumber\\
&+ \sup_{0\le t\le t_p}\left|\left(\frac{Q((1-\eps_1)t-\eps_2)}{Q(t)}\right)^2 - 1\right|\\
&+  \frac{1}{Q(t_p)^2}\left(\prob\left\{\left\|\Delta\right\|_\infty \geq \sigma\eps_2\sqrt{c}\right\} + \prob\left\{\left|\frac{\hsigma}{\sigma}-1\right| \geq \eps_1\right\}\right)\nonumber\\
&\le (1+\eps_1^2+\eps_2^2+\eps_1^2 t_p^4+\eps_2^2 t_p^2 )C(\log p)^{-1-\gamma_1} + C \left(\eps_1^2+\eps_2^2+\eps_1^2 t_p^4+\eps_2^2 t_p^2\right)еее \nonumber\\
&+ \frac{1}{Q(t_p)^2}\left(e^{-c_1 n} + p^{-c_2} + \prob\left\{\left|\frac{\hsigma}{\sigma}-1\right| \geq \eps_1\right\} \right)\,,\label{eq:lengthy}
\end{align}

for some constant $C>0$, where $\gamma_1 = \min(\gamma,1/2)$. In the last inequality above, we applied Lemma~\ref{lem:G1} (Note that $Q(t)\equiv G(t)/2$ by definition). Therefore, by employing bound~\eqref{eq:lengthy} for all $(i,j)\in \Gamma(\gamma,c_0)^c$, we get
\begin{align}
&\frac{1}{p_0^2} \sum_{(i,j)\in \Gamma(\gamma,c_0)^c} \frac{\prob(\tT_i\ge t, \tT_j \ge t)}{Q(t)^2} - 1 \nonumber\\
&\le C \left(\eps_1^2+\eps_2^2+\eps_1^2 t_p^4+\eps_2^2 t_p^2\right) \left(1+(\log p)^{-1-\gamma_1}\right) + C(\log p)^{-1-\gamma_1}\nonumber\\
&\quad+ \frac{1}{Q(t_p)^2}\left(e^{-c_1 n} + p^{-c_2} + \prob\left\{\left|\frac{\hsigma}{\sigma}-1\right| \geq \eps_1\right\} \right) +\frac{|\Gamma(\gamma,c_0)^c|}{p_0^2} -1\,,\label{gamma-Bc}
\end{align}
uniformly for $0\le t\le t_p$, 
and for some positive constants $C, c_1, c_2$. Note that this inequality is obtained by applying .

Combining \eqref{eq:expected-sumoverpairs-a}, \eqref{eq:Psifrag} with bounds~\eqref{gamma-B11}, \eqref{gamma-B12} and~\eqref{gamma-Bc}, we obtain that
\begin{align}
&\E \bigg\{\bigg|\frac{\sum_{i\in\nset}\{\ind(T_i\ge t) - \prob(T_i\ge t)\} }{p Q(t)} \bigg|^2 \bigg\}\nonumber\\
&\le  C \left(\eps_1^2+\eps_2^2+\eps_1^2t_p^4+\eps_2^2t_p^2\right) \left(1+(\log p)^{-1-\gamma_1}+ p^{-1}e^{t^2/2} + p^{\rho-1}e^{(1-\rho)t^2/2}\right) \nonumber\\
&\quad+ \frac{1}{Q(t_p)^2}\left(e^{-c_1 n} + p^{-c_2} + \prob\left\{\left|\frac{\hsigma}{\sigma}-1\right| \geq \eps_1\right\} \right)\nonumber \\
&\quad+ c'p^{\rho-1}e^{(1-\rho)t^2/2} + C p^{-1}e^{t^2/2}+ C(\log p)^{-1-\gamma_1}+ \frac{p^2}{p_0^2} - 1\,, \label{inter-2}
\end{align}
uniformly for $0\le t\le t_p$, some positive constants $C, c_1, c_2$ and for any constant $c'>0$. 

We are now ready to prove the conditions of Lemma~\ref{lem:inter1}, namely Eqs.~\eqref{sup-vpb} and \eqref{int-vpb}. Fix arbitrary constant $\delta>0$. By Chebyshev's inequality, we write
\begin{align}
 &\prob \bigg\{\bigg|\frac{\sum_{i\in \nset}  \ind(T_i\ge t) }{p Q(t)} - 1 \bigg|\ge \delta \bigg\}\nonumber\\
&\le \frac{1}{\delta^2}  \E \bigg\{\bigg|\frac{\sum_{i\in\nset}\{\ind(T_i\ge t) - \prob(T_i\ge t)\} }{p Q(t)} \bigg|^2 \bigg\}\nonumber\\
&\le  \frac{1}{\delta^2} \bigg[C \left(\eps_1^2+\eps_2^2+\eps_1^2t_p^4+\eps_2^2t_p^2\right) \left(1+(\log p)^{-1-\gamma_1}+ p^{-1}e^{t^2/2} + p^{\rho-1}e^{(1-\rho)t^2/2}\right) \nonumber\\
&\quad\quad\quad+ \frac{1}{Q(t_p)^2}\left(e^{-c_1 n} + p^{-c_2} + \prob\left\{\left|\frac{\hsigma}{\sigma}-1\right| \geq \eps_1\right\} \right)\nonumber \\
&\quad\quad\quad+ c'p^{\rho-1}e^{(1-\rho)t^2/2} + C p^{-1}e^{t^2/2}+ C(\log p)^{-1-\gamma_1} + \frac{p^2}{p_0^2} - 1\bigg]\,, \label{eq:inter-3}
\end{align}
where the second step follows from~\eqref{inter-2}, uniformly for $0\le t\le t_p$ and for some constant $C>0$ and an arbitrarily small constant $c'>0$. Hence, by substituting for $t_p = (2\log p - 2\log\log p)^{1/2}$, we obtain
\small
\begin{align}
\label{eq:expression-sup}
&\sup_{0\le t\le t_p} \prob \bigg\{\bigg|\frac{\sum_{i\in \nset}  \ind(T_i\ge t) }{p Q(t)} - 1 \bigg|\ge \delta \bigg\}\nonumber\\
&\le \frac{1}{\delta^2} \bigg[4C \left(\eps_1^2+\eps_2^2+\eps_1^2(\log p)^2+\eps_2^2\log p\right)\left(1+(\log p)^{-1-\gamma_1}
+ \left(\log p\right)^{-1} + \left(\log p\right)^{-1+\rho}\right) \nonumber\\
&\quad\quad\quad+ p^2 \left(e^{-c_1 n} + p^{-c_2} + \prob\left\{\left|\frac{\hsigma}{\sigma}-1\right| \geq \eps_1\right\} \right) \nonumber\\
&\quad\quad\quad+ C\left(\log p\right)^{-1} + c'\left(\log p\right)^{-(1-\rho)} + C(\log p)^{-1-\gamma_1}+ \frac{p^2}{p_0^2} - 1\bigg]\,.
\end{align}
\normalsize
Recall that $\eps_2 = s_0(\log p)/\sqrt{n}$. We take $\eps_1 = \sqrt{s_0(\log p)/n}$. By~\cite[Lemma 3.3]{javanmard2014confidence}, we have that for this choice of $\eps_1$, 
$\prob\left\{\left|\hsigma/\sigma-1\right|\geq \eps_1\right\}\to 0$ and hence Eq.\eqref{sup-vpb} holds.

 Likewise,~\eqref{int-vpb} holds because continuing from~\eqref{eq:inter-3} and by applying reverse Fatou Lemma, we can write
 
 \begin{align}
 \label{eq:expression-int}
 &\int_{0}^{t_p}\prob \bigg\{\bigg|\frac{\sum_{i\in\nset}  \ind(T_i \ge t) }{p G(t)} - 1 \bigg|\ge \delta \bigg\} \de t \le\nonumber \\
&\int_{0}^{t_p} \Big[C \left(\eps_1^2+\eps_2^2+\eps_1^2t_p^4+\eps_2^2t_p^2\right) \left(1+(\log p)^{-1-\gamma_1}+ p^{-1}e^{t^2/2} + p^{\rho-1}e^{(1-\rho)t^2/2}\right) \nonumber\\
&\quad+ t_p^2 e^{t_p^2}\left(e^{-c_1 n} + p^{-c_2} + \prob\left\{\left|\frac{\hsigma}{\sigma}-1\right| \geq \eps_1\right\} \right)\\
&\quad+ c'p^{\rho-1}e^{(1-\rho)t^2/2} + C p^{-1}e^{t^2/2}+ C(\log p)^{-1-\gamma_1}\Big]\, \de t\nonumber \\
&\leq C \left(\eps_1^2t_p+\eps_2^2t_p+\eps_1^2t_p^5+\eps_2^2t_p^3\right) \left(1+(\log p)^{-1-\gamma_1} + p^{-1}e^{t^2/2}+p^{\rho-1}e^{(1-\rho)t^2/2}\right)\nonumber\\
&+ t_p^3 e^{t_p^2}\left(e^{-c_1 n} + p^{-c_2} + \prob\left\{\left|\frac{\hsigma}{\sigma}-1\right| \geq \eps_1\right\} \right) \\
&+c'p^{\rho-1}t_pe^{(1-\rho)t_p^2/2} + C p^{-1}t_pe^{t_p^2/2} + Ct_p(\log p)^{-1-\gamma_1}\nonumber\\
&\leq 2C \left(\eps_1^2(\log p)^{5/2}+\eps_2^2(\log p)^{3/2}\right) \left(1+(\log p)^{-1-\gamma_1} + \left(\log p\right)^{-1} + \left(\log p\right)^{-1+\rho}\right) \nonumber\\ 
&+p^2(\log p)^{-1/2}\left(e^{-c_1 n} + p^{-c_2} + \prob\left\{\left|\frac{\hsigma}{\sigma}-1\right| \geq \eps_1\right\}\right)\\
&+ c (\log p)^{-(1/2-\rho)} + C (\log p)^{-1/2} + C(\log p)^{-1/2-\gamma_1}\nonumber\\
&= o((\log p)^{-1/2}) = o(v_p)\,.
 \end{align}

In the last step we used the probabilistic bound on $|\hsigma/\sigma-1|$, given in~\cite[Theorem 2.1]{SZ-scaledLasso}, with $\eps_1 = \sqrt{s_0(\log p)/n}$, and assumption $s_0 = o\left(\sqrt{n}/ (\log p)^{2}\right)$. This shows that Eq.~\eqref{int-vpb} holds and hence completes the proof.

\subsection{Proof of Theorem~\ref{thm:power}}\label{proof:thm-power}

The threshold $t_0$ retuned by the FCD procedure is data-dependent. To analyze the power, we first upper bound $t_0$ by a data-independent threshold $t_*$.
\begin{lemma}\label{lem:tstar}
Under the assumptions of Theorem~\ref{thm:power}, we have
\[t_0\le t_*\,,\quad t_* = \Phi^{-1}\left(1-\frac{qs_0}{2p}(1-o(1))\right)\,.\]
\end{lemma}
Proof of Lemma~\ref{lem:tstar} is given in Appendix~\ref{proof:lem-tstar}.

Since $t_0 \le t_*$ by Lemma~\ref{lem:tstar}, if we replace $t_0$ by $t_*$, we obtain a lower bound on the power. 
For fixed arbitrarily small constants $c_0$, $\delta$, $\eps$, define 
\[
\cG = \cG(\delta,c_0,\eps) = \Big\{\max|\Lambda_{ii}-\Omega_{ii}| \le c_0,\; |\hsigma/\sigma-1|\le \delta,\; \|\Delta\|_\infty \le \eps \Big\}\,.
\]
Define $S_+ \equiv \{i\in [p]: \theta_{0,i} > 0\}$ and $S_- \equiv \{i\in [p]: \theta_{0,i} < 0\}$. Therefore, $S = S_+\cup S_-$. We have

\begin{align}
\power &= \E\bigg[\frac{|\{j\in \hS:\, \hsign_j = \sign(\theta_{0,j})\}|}{\max(|S|,1)} \bigg]\nonumber\\
&=\frac{1}{s_0}\sum_{i\in S_+} \prob(T_i\ge t_*)+ \frac{1}{s_0}\sum_{i\in S_0^+} \prob(T_i \le -t_*)\nonumber\\
&=\frac{1}{s_0}\sum_{i\in S_+} \prob\Big(\frac{\sqrt{n}\dth_i}{\hsigma \sqrt{\Lambda_{ii}}}\ge t_*\Big)
+\frac{1}{s_0}\sum_{i\in S_-} \prob\Big(\frac{\sqrt{n} \dth_i}{\hsigma \sqrt{\Lambda_{ii}}}\le -t_*\Big)\nonumber\\
& = \frac{1}{s_0} \sum_{i\in S_+} \prob\left(\frac{\sigma}{\hsigma} \tZ_i + \frac{\sqrt{n}\theta_{0,i} + \Delta_i}{\hsigma \sqrt{\Lambda_{ii}}}   \ge t_*\right)\\
&+ \frac{1}{s_0}\sum_{i\in S_-} \prob\left(\frac{\sigma}{\hsigma} \tZ_i + \frac{\sqrt{n}\theta_{0,i} + \Delta_i}{\hsigma \sqrt{\Lambda_{ii}}}   \le -t_*\right)
\label{eq:powerLB1}
\end{align}

Define $\eta_i\equiv (\sqrt{n}\theta_{0,i}+\Delta_i)/(\sigma \sqrt{\Lambda_{ii}})$. On event $\cG$, we have 
\[\eta_i \ge \eta_{i}^-\equiv \frac{\sqrt{n}\theta_{0,i} -\eps}{\sigma\sqrt{\Omega_{ii}+c_0}}\,,\quad\quad
\eta_i \le \eta_{i}^+\equiv \frac{\sqrt{n}\theta_{0,i} +\eps}{\sigma\sqrt{\Omega_{ii}-c_0}}\,. \]
Using Equation~\eqref{eq:powerLB1}, we have

\begin{align}
\power& \ge \frac{1}{s_0}\sum_{i\in S_+} \prob\left( \Big[ Z_i + \eta_i   \ge \frac{\hsigma}{\sigma}  t_*\Big] \cdot \ind(\cG)\right)\\
&+\frac{1}{s_0}\sum_{i\in S_-} \prob\left( \Big[ Z_i + \eta_i   \le \frac{-\hsigma}{\sigma}  t_*\Big] \cdot \ind(\cG)\right)  - \prob(\cG^c)\nonumber\\
&\ge \frac{1}{s_0}\sum_{i\in S_+} \prob\left(\Big[Z_i + \eta_{i}^-   \ge (1+\delta)  t_*\Big] \cdot \ind(\cG)\right)\\
&+\frac{1}{s_0}\sum_{i\in S_-} \prob\left(\Big[Z_i + \eta_{i}^+   \le -(1+\delta)  t_*\Big] \cdot \ind(\cG)\right) -\prob(\cG^c)\nonumber\\
&= \frac{1}{s_0}\sum_{i\in S_+} \prob\left(Z_i + \eta_{i}^-   \ge (1+\delta)  t_*\right) \prob(\cG)\\
&+\frac{1}{s_0}\sum_{i\in S_-} \prob\left(Z_i + \eta_{i}^+   \le -(1+\delta)  t_*\right) \prob(\cG)-\prob(\cG^c)\,.\nonumber
\end{align}

Recall that $s_0=o(\sqrt{n}/(\log p)^2)$ as per Condition $(i)$, and by using Proposition~\ref{propo:biasvardecompose} and lemma~\ref{lemma:sigmahatoversigma}, event $\cG$ holds with high probability and indeed it is easy to see that for $\theta_{\min} >(\sigma/\sqrt{n}) \sqrt{2\log(p/s_0)}$, we have 
\[\underset{n\to\infty}{\lim\sup} \frac{\prob(\cG^c)}{1-\beta(\theta_0,n)} = 0\,. \]
Therefore,

\begin{align}
&\underset{n\to\infty}{\lim\inf} \frac{\power}{1-\beta(\theta_0,n)}  \ge\nonumber\\
&\underset{n\to\infty}{\lim\inf} 
\frac{1}{s_0(1-\beta(\theta_0,n))}\bigg[\sum_{i\in S_+} \prob\left(Z_i + \eta_{i}^-   \ge (1+\delta)  t_*\right)\\
&\hspace{1.3in}+\sum_{i\in S_-} \prob\left(Z_i + \eta_{i}^+   \le -(1+\delta)  t_*\right)\bigg]\,.
\end{align}

Since the above bound holds for all $\eps$, $\delta$, $c_0 > 0$, we get

\begin{align}
&\underset{n\to\infty}{\lim\inf} \frac{\power}{1-\beta(\theta_0,n)}  \nonumber\\
&\ge \underset{n\to\infty}{\lim\inf} 
\frac{1}{s_0(1-\beta(\theta_0,n))}\bigg[\sum_{i\in S_+} \prob\left(Z_i + \frac{\sqrt{n}\theta_{0,i}}{\sigma \sqrt{\Omega_{ii}}}  \ge   t_*\right)\\
&\hspace{1.5in}+\sum_{i\in S_-} \prob\left(Z_i + \frac{\sqrt{n}\theta_{0,i}}{\sigma \sqrt{\Omega_{ii}}}   \le -  t_*\right)\bigg]\nonumber\\
&=\underset{n\to\infty}{\lim\inf} \frac{1}{s_0(1-\beta(\theta_0,n))} 
\Big\{\sum_{i\in S} \left(1-\Phi\left(t_* -  \frac{\sqrt{n}|\theta_{0,i}|}{\sigma \sqrt{\Omega_{ii}}}\right)\right)  \Big\} \nonumber\\
&=\underset{n\to\infty}{\lim\inf} \frac{1}{(1-\beta(\theta_0,n))} 
\Big\{ \frac{1}{s_0}\sum_{i\in S} F\left(\frac{qs_0}{p},\frac{\sqrt{n}|\theta_{0,i}|}{\sigma \sqrt{\Omega_{ii}}} \right)  \Big\} = 1\,.
\end{align}

The last step holds by using the definition of function $F(\cdot,\cdot)$, given by Equation~\eqref{def:G}, and the fact that $Z_i|X\sim\normal(0,1)$.

\subsection{Proof of Theorem~\ref{thm:main2}}
\label{proof:thm-main2}
The proof follows the proof of Theorem~\ref{thm:main}. Note that for the results of theorem 
to hold, it suffices that the conditions of Lemma~\ref{lem:inter1} to be satisfied. 
The result in ~\cite[Theorem 3.8]{javanmard2015biasing}, implies that under the conditions 
of Theorem~\ref{thm:main2}, for some constants $C$, $c$, and $n \geq \max(25\log p, cs_0\log(p/s_0))$, we have
\begin{align}
\label{eq:biasboundgaussian2}
\P\left(\left\|\Delta\right\|_\infty\geq C\tau\sigma\sqrt{\frac{s_0}{n}}\log p\right) \leq 2pe^{-C_{\min}n/(16s_0)} + pe^{-n/1000} + 8p^{-1}.
\end{align}
Using this, under the assumptions of Theorem~\ref{thm:main2}, letting
$\eps_2 = (\log p) \tau_0\sqrt{s_0/n}$, and following the same steps as in 
the proof of Theorem~\ref{thm:main}, we will reach the following equation which is similar 
to Eq.~\eqref{eq:expression-sup}
\small
\begin{align}
\label{eq:expression-sup2}
&\sup_{0\le t\le t_p} \prob \bigg\{\bigg|\frac{\sum_{i\in \nset}  \ind(T_i\ge t) }{p Q(t)} - 1 \bigg|\ge \delta \bigg\}\nonumber\\
&\le \frac{1}{\delta^2} \bigg[4C \left(\eps_1^2+\eps_2^2+\eps_1^2(\log p)^2+\eps_2^2\log p\right) \left(1+(\log p)^{-1-\gamma_1}
+ \left(\log p\right)^{-1} + \left(\log p\right)^{-1+\rho}\right) \nonumber\\
&\quad\quad\quad+ p^2 \left(2pe^{-C_{\min}n/(16s_0)} + pe^{-n/1000} + 8p^{-1} + \prob\left\{\left|\frac{\hsigma}{\sigma}-1\right| \geq \eps_1\right\} \right) \nonumber\\
&\quad\quad\quad+ C\left(\log p\right)^{-1} + c'\left(\log p\right)^{-1+\rho} + C(\log p)^{-1-\gamma_1}+ \frac{p^2}{p_0^2} - 1\bigg]\,.
\end{align}
\normalsize
By taking $\eps_1 = \sqrt{s_0(\log p)/n}$ in Eq.~\eqref{eq:expression-sup} and
replacing $\eps_2 = (\log p) \tau_0\sqrt{s_0/n}$,
we deduce that Eq.~\eqref{sup-vpb}
holds. 
Similarly, using Eq.~\eqref{eq:biasboundgaussian2}, we reach 
\small
\begin{align}
\label{eq:expression-int2}
 &\int_{0}^{t_p}\prob \bigg\{\bigg|\frac{\sum_{i\in\nset}  \ind(T_i \ge t) }{p G(t)} - 1 \bigg|\ge \delta \bigg\} \de t \nonumber \\
&\leq 2C \left(\eps_1^2(\log p)^{5/2}+\eps_2^2(\log p)^{3/2}\right) \left(1+(\log p)^{-1-\gamma_1} + \left(\log p\right)^{-1} + \left(\log p\right)^{-(1-\rho)}\right) \nonumber\\ 
&+p^2(\log p)^{-1/2}\left(2pe^{-C_{\min}n/(16s_0)} + pe^{-n/1000} + 8p^{-1}  + \prob\left\{\left|\frac{\hsigma}{\sigma}-1\right| \geq \eps_1\right\}\right) \nonumber\\ 
&+ c (\log p)^{-(1/2-\rho)} + C (\log p)^{-1/2} + C(\log p)^{-1/2-\gamma_1}.
\end{align}
\normalsize
which is similar to Eq.~\eqref{eq:expression-int}. 
Again, by taking $\eps_1 = \sqrt{s_0(\log p)/n}$ and $\eps_2 = (\log p) \tau_0\sqrt{s_0/n}$ we deduce
that Eq.~\eqref{int-vpb} holds too. Hence, the desired results hold under the conditions 
of the Theorem.

\subsection{Proof of Theorem~\ref{thm:main3}}
\label{proof:thm-main3}
The proof is similar to the proof of Theorem~\ref{thm:main2}. Here, using the result in ~\cite[Theorem 3.13]{javanmard2015biasing}, under the conditions 
of Theorem~\ref{thm:main3}, for some constants $C$, $c$, and $n \geq s_0\log p$, we have
\begin{align}
\label{eq:biasboundgaussian3}
\P\left(\left\|\Delta\right\|_\infty\geq C\tau\sigma\sqrt{\frac{s_0}{n}}\log p + C\sigma\min(s_0, s_{\Omega})\frac{\log p}{\sqrt{n}}\right) &\leq 2pe^{-C_{\min}n/(16s_0)} \\
&+ pe^{-cn} + 8p^{-1}.
\end{align}
Here, by taking
$\eps_2 = (\log p) \tau_0\sqrt{s_0/n} + \min(s_0, s_{\Omega})\log p/\sqrt{n}$,
 we will reach the following equation which is similar 
to Eqs.~\eqref{eq:expression-sup},~\eqref{eq:expression-sup2}
\small
\begin{align}
&\sup_{0\le t\le t_p} \prob \bigg\{\bigg|\frac{\sum_{i\in \nset}  \ind(T_i\ge t) }{p Q(t)} - 1 \bigg|\ge \delta \bigg\}\nonumber\\
&\le \frac{1}{\delta^2} \bigg[4C \left(\eps_1^2+\eps_2^2+\eps_1^2(\log p)^2+\eps_2^2\log p\right) \left(1+(\log p)^{-1-\gamma_1}
+ \left(\log p\right)^{-1} + \left(\log p\right)^{-1+\rho}\right) \nonumber\\
&\quad\quad\quad+ p^2 \left(2pe^{-C_{\min}n/(16s_0)} + pe^{-cn} + 8p^{-1} + \prob\left\{\left|\frac{\hsigma}{\sigma}-1\right| \geq \eps_1\right\} \right) \nonumber\\
&\quad\quad\quad+ C\left(\log p\right)^{-1} + c'\left(\log p\right)^{-(1-\rho)} + C(\log p)^{-1-\gamma_1}+ \frac{p^2}{p_0^2} - 1\bigg]\,\nonumber.
\end{align}
\normalsize
By taking $\eps_1 = \sqrt{s_0(\log p)/n}$ in Eq.~\eqref{eq:expression-sup} and
replacing $\eps_2 = (\log p) \tau_0\sqrt{s_0/n} + \min(s_0, s_{\Omega})\log p/\sqrt{n}$,
we deduce that Eq.~\eqref{sup-vpb}
holds. 
Similarly, using Eq.~\eqref{eq:biasboundgaussian3}, we reach
\small
\begin{align}
 &\int_{0}^{t_p}\prob \bigg\{\bigg|\frac{\sum_{i\in\nset}  \ind(T_i \ge t) }{p G(t)} - 1 \bigg|\ge \delta \bigg\} \de t \nonumber \\
&\leq 2C \left(\eps_1^2(\log p)^{5/2}+\eps_2^2(\log p)^{3/2}\right) \left(1+(\log p)^{-1-\gamma_1} + \left(\log p\right)^{-1} + \left(\log p\right)^{-(1-\rho)}\right) \nonumber\\ 
&+p^2(\log p)^{-1/2}\left(2pe^{-C_{\min}n/(16s_0)} + pe^{-cn} + 8p^{-1}  + \prob\left\{\left|\frac{\hsigma}{\sigma}-1\right| \geq \eps_1\right\}\right) \nonumber\\ 
&+ c (\log p)^{-(1/2-\rho)} + C (\log p)^{-1/2} + C(\log p)^{-1/2-\gamma_1}.\nonumber 
\end{align}
\normalsize
which is similar to Eqs.~\eqref{eq:expression-int},~\eqref{eq:expression-int2}. 
Again, by taking $\eps_1 = \sqrt{s_0(\log p)/n}$, $\eps_2 = (\log p) \tau_0\sqrt{s_0/n} + \min(s_0, s_{\Omega})\log p/\sqrt{n}$ we deduce
that Eq.~\eqref{int-vpb} holds too. Hence, the desired results hold under the conditions 
of the Theorem.

\section*{Acknowledgements}

A. Javanmard was partially supported by the NSF CAREER Award 1844481 and a Google Faculty Research Award. A. Javanmard would also like
to acknowledge the financial support of the Office of the Provost at the University of Southern
California through the Zumberge Fund Individual Grant Program. 


\newpage
\appendix
\section{Proof of Technical Lemmas}
\subsection{Proof of Lemma~\ref{lem:G2}}\label{proof:lem-G2}
For $t\ge 0$, we write 
\begin{align}\label{frac-1}
\frac{G\left((1- \delta)t - \eps\right)}{G(t)} = 1+ \frac{\int_{(1-\delta)t - \eps}^{t}\phi(x)\de x}{G(t)}
\le
 1 + \frac{(\eps + \delta t)\phi((1-\delta)t-\eps)}{G(t)}\,,
\end{align}
where we used that $\phi(t)$ is a decreasing function.  We next separate the cases of $t\in (0,1)$ and $t\ge 1$.

For $0<t<1$, we use the following bound
\begin{align}\label{frac-0}
\phi(t)\le (\sqrt{4+t^2}-t) \phi(t) \le G(t)\,,
\end{align}
where the last step is due to Birnbaum~\cite{birnbaum1942inequality}.

Moreover, for all $t\ge 0$,
\begin{align}
\frac{\phi((1-\delta)t-\eps)}{\phi(t)} &= \exp\Big\{t(\delta t+\eps)- \frac{1}{2}((1-\delta)t+\eps)^2\Big\}\nonumber\\
&\le \exp\Big\{t(\delta t+\eps)\Big\}\le e^2\,,\label{frac-2}
\end{align}
because by our assumption $\delta^2t\le 1$ and $\eps t\le 1$.

By employing Eqs.~\eqref{frac-0} and~\eqref{frac-2} into Eq.~\eqref{frac-1}, we obtain
\begin{align}\label{t-small}
\frac{G\left((1- \delta)t - \eps\right)}{G(t)}\le 1+ e^2(\eps+\delta t)\le 1 + e^2(\eps+\delta)\,.
\end{align}
For $t\ge 1$, using Lemma~\ref{lem:G1}, we have that $G(t) \ge \phi(t)/t$ and hence by using Eq.~\eqref{frac-2} into Eq.~\eqref{frac-1}, we get
\begin{align}\label{t-large}
\frac{G\left((1- \delta)t - \eps\right)}{G(t)}\le 1+ e^2 t(\eps+\delta t)\,.
\end{align}
The result follows by combining the bound~\eqref{t-small} and~\eqref{t-large}.

\subsection{Proof of Lemma~\ref{lem:inter1}}\label{proof:lem-inter1}
We write

\begin{align*}
\prob&\bigg[ \max_{0\le j\le b}\, \bigg|\frac{\sum_{i\in\nset} \{\ind(T_i \ge \tau_j) - Q(\tau_j)\} }{p Q(\tau_j)} \bigg|\ge \delta \bigg] \\
&\hspace{1in}\le
\sum_{j=1}^b \prob\bigg[ \bigg|\frac{\sum_{i\in\nset} \{\ind(T_i \ge \tau_j) - Q(\tau_j)\} }{p Q(\tau_j)} \bigg|\ge \delta \bigg]\\
&\hspace{1in}\le \frac{1}{v_p} \int_{0}^{t_p} \prob\bigg\{\bigg|\frac{\sum_{i\in\nset} \ind(T_i\ge t)}{pQ(t)} - 1 \bigg|\ge \delta \bigg\} \de t\\
&\hspace{1in}+
\sum_{j=b-1}^b \prob\bigg[ \bigg|\frac{\sum_{i\in\nset} \{\ind(T_i \ge \tau_j) - Q(\tau_j)\} }{p Q(\tau_j)} \bigg|\ge \delta \bigg]
\end{align*}

Therefore, it suffices to show that
\[
\int_{0}^{t_p} \prob\bigg\{\bigg|\frac{\sum_{i\in\nset} \ind(T_i\ge t)}{pQ(t)} - 1 \bigg|\ge \delta \bigg\} \de t = o(v_p)\,,
\]
and
\[
\sup_{0\le t\le t_p} \prob\bigg\{\bigg|\frac{\sum_{i\in\nset} \ind(T_i\ge t)}{pQ(t)} - 1 \bigg|\ge \delta \bigg\} \de t = o(1)\,,
\]
which are the conditions of the lemma.

\subsection{Proof of Lemma~\ref{lem:tstar}}\label{proof:lem-tstar}
We first show that $t_*< \sqrt{2\log (p/s_0)}$. Assuming otherwise, we have $G(t_*)<G(\sqrt{2\log (p/s_0)})$ because $G(t)$ is decreasing. By definition of $t_*$, and Lemma~\ref{lem:G1} this results in
\begin{align}
\frac{qs_0}{p}(1-o(1)) = G(t_*) < G(\sqrt{2\log (p/s_0)}) &\le \sqrt{\frac{2}{\pi}} \frac{e^{-\log (p/s_0)}}{\sqrt{2\log (p/s_0)}} \\
&= \frac{s_0}{p\sqrt{\pi \log (p/s_0)}  }\,, 
\end{align}
which is a contradiction.

Now, given that $t_*<\sqrt{2\log(p/s_0)} < \sqrt{2\log p}$, if the claim is not true, by definition of $t_0$, we should have
\begin{align}\label{power0}
\frac{2p(1-\Phi(t_*))}{R(t_*)\vee 1} > q\,.
\end{align}
We next show that $R(t_*) \ge s_0(1-o(1))$. 

Define 
\[
\cG = \cG(\delta,c_0,\eps) = \Big\{\max|\Lambda_{ii}-\Omega_{ii}| \le c_0,\, |\hsigma/\sigma-1|\le \delta,\, \|\Delta\|_\infty \le \eps \Big\}
\]

Define $\hS(t_*) = \{ i\in[p]:\, |T_i|> t_*\}$. Using Proposition~\eqref{propo:biasvardecompose}, for fixed $i\in S$, we have
\begin{align}
\prob(i\notin \hS(t_*)) &= \prob(|T_i|\le t_*)\nonumber\\
&= \prob\left(\Big|\frac{\sqrt{n} \theta_{0,i}}{\hsigma \sqrt{\Lambda_{ii}}}+\frac{\sigma}{\hsigma}Z_i+ \frac{\Delta_i}{\hsigma\sqrt{\Lambda_{ii}}}\Big|\le t_* \right)\,,\label{power1}
\end{align} 
with $Z_i\sim\normal(0,1)$. Define $\eta_i\equiv (\sqrt{n}\theta_{0,i}+\Delta_i)/(\sigma \sqrt{\Lambda_{ii}})$. On event $\cG$, we have 
\[|\eta_i| \ge \eta_{i,*}\equiv \frac{\sqrt{n}|\theta_{0,i}| -\eps}{\sigma\sqrt{\Omega_{ii}+c_0}}\]

Continuing from Equation~\eqref{power1}, we have
\begin{align*}
\prob(i\notin \hS(t_*)) &= \prob\left(|Z_i+ \eta_i|\le \frac{\hsigma}{\sigma} t_* \right)\le \prob\left(\left[|Z_i+ \eta_{i,*}|\le \frac{\hsigma}{\sigma} t_*\right]\cdot \ind(\cG) \right) + \prob(\cG^c)\\
&\le \prob\left(\left[ \eta_{i,*} -\frac{\hsigma}{\sigma} t_*\le |Z_i|\right]\cdot \ind(\cG) \right) + \prob(\cG^c)\,.
\end{align*}
Given that $\theta_{0,i} > ({\sigma}/\sqrt{n}) \sqrt{2\Omega_{ii}\log(p/s_0)}$ and $t_*<\sqrt{2\log(p/s_0)}$, we can choose $\delta$, $c_0$, $\eps$ and $\eps_0$ small enough such that on event $\cG = \cG(\delta,c_0,\eps)$, 
\[
\eta_{i,*}- \frac{\hsigma}{\sigma} t_* \ge t_*\,,
\]
and therefore
\begin{align}
\prob(i\notin \hS(t_*))&\le  \prob\left(\left[ \eta_{i,*} -\frac{\hsigma}{\sigma} t_*\le |Z_i|\right]\cdot \ind(\cG) \right) + \prob(\cG^c)\nonumber\\
&\le \prob\left(\left( t_*\le |Z_i|\right)\cdot \ind(\cG) \right) + \prob(\cG^c)\nonumber\\
&\le G(t_*)+ \prob(\cG^c)\nonumber\\
&\le \left(\frac{qs_0}{p}\right) + \prob(\cG^c)\label{power2}
\end{align}
Since $\prob(\cG^c)\to 0$ and $s_0 = o(\sqrt{n}/(\log p)^2)$, we can choose a deterministic sequence $L_n \to \infty$, arbitrarily slow, as $n\to \infty$, such that 
$L_n\prob(\cG^c)\to 0$ and $L_n (s_0/p) \to 0$. Letting $A_n \equiv  (qs_0/p) + \prob(\cG^c)$, we have $L_nA_n\to 0$.  

By applying Markov inequality, we obtain
\begin{align}
\prob(|S\cap \hS(t_*)^c|\ge s_0L_n A_n) &\le \frac{1}{s_0L_n A_n} \E(|S_0\cap \hS(t_*)^c|)\nonumber\\
&\le \frac{s_0 A_n}{s_0L_n A_n} = \frac{1}{L_n}\,,
\end{align}
where the last inequality follows from~\eqref{power2}. Therefore, with high probability, $|S_0\cap \hS(t_*)^c|\le s_0 L_n A_n$, which implies that
\begin{align}\label{power3}
R(t_*) = |\hS(t_*)| \ge |S| -|S\cap \hS(t_*)^c| \ge s_0 (1- L_n A_n) \,,
\end{align}
as claimed.

Now using Equation~\eqref{power3} in Equation~\eqref{power0}, we arrive at
\[
1-\Phi(t_*) > \frac{qs_0}{2p}(1-L_n A_n)\,. 
\]
Therefore, for $t_*$, given by
\[
t_* = \Phi^{-1}\left(1 - \frac{qs_0}{2p} (1-2L_n A_n)\right)\,,
\]
we reach a contradiction which proves our claim $t_0\le t_*$ is correct. The proof is complete by noting that $L_n A_n = o(1)$ by choice of sequence $L_n$.
\subsection{Proof of Corollary~\ref{coro:unit-power}}\label{proof:unit-power}
Define
\[\alpha_n = \frac{qs_0}{p}\,,\quad u_n\equiv \frac{\sqrt{n}\theta_{\min}}{\sigma\sqrt{\Omega_{ii}}}\,. \]
Using Corollary~\ref{coro:theta-min-power}, it suffices to show that $F\left(\alpha_n,u_n\right) = 1 - \Phi(\Phi^{-1}(1-\alpha_n/2) - u_n) \to 1$. Equivalently, we show that $\Phi^{-1}(1-\alpha_n/2) - u_n \to -\infty$.

By Lemma~\ref{lem:G1}, we have
\begin{align}
G(\sqrt{2\log(2/\alpha_n)}) < \frac{2\phi(\sqrt{2\log(2/\alpha_n)})}{\sqrt{2\log(2/\alpha_n)}} < 2\phi(\sqrt{2\log(2/\alpha_n)}) = \alpha_n\,.
\end{align}
Since $G$ is a decreasing function, by applying $G^{-1}$ on both sides, we get
\[
\Phi^{-1}(1-{\alpha}/{2}) = G^{-1}(\alpha_n)\le \sqrt{2\log(2/\alpha_n)}
\]
Using the above bound, we have
\begin{align}\label{eq:un-LB}
u_n - \Phi^{-1}(1-\alpha_n/2) > u_n - \sqrt{2\log(2/\alpha_n)}
\end{align}
By the assumption on $\theta_{\min}$, we have that the left-hand side of~\eqref{eq:un-LB} goes to $\infty$, which completes the proof.

\bibliographystyle{Chicago}

\newcommand{\etalchar}[1]{$^{#1}$}
\providecommand{\bysame}{\leavevmode\hbox to3em{\hrulefill}\thinspace}
\providecommand{\MR}{\relax\ifhmode\unskip\space\fi MR }
\providecommand{\MRhref}[2]{%
  \href{http://www.ams.org/mathscinet-getitem?mr=#1}{#2}
}
\providecommand{\href}[2]{#2}

\end{document}